\pdfoutput=1 
\documentclass[a4paper,11pt,table]{article}

\usepackage{jheppub} 

\usepackage[T1]{fontenc} 
\usepackage[utf8]{inputenc} 
\usepackage{microtype} 
\usepackage{lmodern} 
\usepackage{booktabs} 
\usepackage{mdframed} 
\usepackage{graphicx}
\usepackage{todonotes} 
\usepackage{slashed}
\usepackage[utf8]{inputenc}
\usepackage{cancel}
\usepackage{tikz}
\usetikzlibrary{decorations.pathmorphing,backgrounds,calc,shapes,arrows,shadows}
\tikzset{zigzag/.style={decorate,decoration=zigzag}}
\tikzset{
    partial ellipse/.style args={#1:#2:#3}{
        insert path={+ (#1:#3) arc (#1:#2:#3)}
    }
}

\usepackage{pdflscape} 
\usepackage{multirow} 

\usepackage{caption}
\usepackage{array} 
\newcolumntype{L}{>{$}l<{$}}
\newcolumntype{C}{>{$}c<{$}}

\usepackage{hyperref}
 \usepackage{color}
 \definecolor{dark-red}{rgb}{0.4,0.15,0.15}
 \definecolor{dark-blue}{rgb}{0.15,0.15,0.4}
 \definecolor{medium-blue}{rgb}{0,0,0.5}

 \hypersetup{colorlinks, linkcolor={dark-red}, citecolor={dark-blue}, urlcolor={medium-blue}}

\usepackage{todonotes} 

\usepackage{amsmath,amsfonts,amssymb}
\usepackage{eucal} 
\usepackage{mleftright} \mleftright
\usepackage{tensor} 
\usepackage{braket} 
\usepackage{mathrsfs} 
\usepackage{array}

\providecommand*{\dd}{\mathop{}\!d}
\renewcommand*{\dd}{\mathop{}\!d}

\providecommand*{\vd}{\mathop{}\!\delta}
\renewcommand*{\vd}{\mathop{}\!\delta}

\usepackage{mathrsfs} 
\providecommand*{\lagr}{\mathscr{L}}
\renewcommand*{\lagr}{\mathscr{L}}

\providecommand*{\R}{{\mathbb{R}}}
\renewcommand*{\R}{{\mathbb{R}}}

\DeclareMathOperator{\tr}{tr}

\usepackage{tikz}
\usepackage{tikz-cd}
\usepackage{tikz-3dplot}

\newcommand{\shi}{5}

\newcommand\parmp{\mathbin{\vcenter{\hbox{%
  \oalign{$\scriptstyle({-})$\cr
          \noalign{\kern-0.8ex}
          \hfil$\scriptscriptstyle+$\hfil\cr}%
}}}}

\providecommand{\Bt}{{\mathtt{B}}}
\renewcommand{\Bt}{{\mathtt{B}}}

\providecommand{\Ht}{{\mathtt{H}}}
\renewcommand{\Ht}{{\mathtt{H}}}

\providecommand{\Pt}{{\mathtt{P}}}
\renewcommand{\Pt}{{\mathtt{P}}}
\providecommand{\Lt}{{\mathtt{L}}}
\renewcommand{\Lt}{{\mathtt{L}}}

\providecommand{\Zt}{{\mathtt{Z}}}
\renewcommand{\Zt}{{\mathtt{Z}}}
\providecommand{\Mt}{{\mathtt{M}}}
\renewcommand{\Mt}{{\mathtt{M}}}

\providecommand{\Et}{{\mathtt{E}}}
\renewcommand{\Et}{{\mathtt{E}}}

\providecommand{\Mt}{{\mathtt{M}}}
\renewcommand{\Mt}{{\mathtt{M}}}

\providecommand{\ett}{{\mathtt{e}}}
\renewcommand{\ett}{{\mathtt{e}}}
\providecommand{\Dt}{{\mathtt{D}}}
\renewcommand{\Dt}{{\mathtt{D}}}
\providecommand{\Ct}{{\mathtt{C}}}
\renewcommand{\Ct}{{\mathtt{C}}}

\newcommand{\extd}{\dd}

\newcommand{\lc}{\hat{\Lambda}}
\newcommand{\cc}{\hat{c}}
\newcommand{\tc}{\hat{C}}

\usepackage{pifont}
\newcommand{\cmark}{{\LARGE \ding{51}}}

\newcommand{\yes}{{\text{\ding{51}}}}
\newcommand{\no}{{\text{\ding{55}}}}

\newcommand{\slR}{\mathfrak{sl}(2,\R)}
\newcommand{\so}{\mathfrak{so}(3)}
\newcommand{\So}{\mathfrak{so}}
\newcommand{\poi}{\text{VI}_{0}}
\newcommand{\poic}{\text{VI}^{c}_{0}}
\newcommand{\eu}{\text{VII}_{0}}
\newcommand{\euc}{\text{VII}^{c}_{0}}

\newcommand{\hei}{\text{II}}

\newcommand{\heicc}{\text{II}^{cc}}

\newcommand{\timet}{t}

\usepackage{amsthm} 
\theoremstyle{plain}

\theoremstyle{definition}

\theoremstyle{remark}

\usepackage{float}

\title{\boldmath Limits of JT gravity}

\author[a,1]{Daniel Grumiller\note{\href{https://orcid.org/0000-0001-7980-5394}{ORCID: 0000-0001-7980-5394}},}
\author[b,2]{Jelle Hartong\note{\href{https://orcid.org/0000-0003-0498-0029}{ORCID: 0000-0003-0498-0029}},}
\author[b,3]{Stefan Prohazka\note{\href{https://orcid.org/0000-0002-3925-3983}{ORCID: 0000-0002-3925-3983}}}
\author[c,4]{and Jakob Salzer\note{\href{https://orcid.org/0000-0002-9560-344X}{ORCID: 0000-0002-9560-344X}}}

\affiliation[a]{Institute for Theoretical Physics, TU Wien, \\
  Wiedner Hauptstrasse 8–10/136, A-1040 Vienna, Austria}
\affiliation[b]{School of Mathematics and Maxwell Institute for
  Mathematical Sciences, University of Edinburgh,\\
  Peter Guthrie Tait Road, Edinburgh EH9 3FD, UK }
\affiliation[c]{Center for the Fundamental Laws of Nature, Harvard
  University, \\
  Cambridge, MA 02138, USA}

\emailAdd{grumil@hep.itp.tuwien.ac.at}
\emailAdd{jelle.hartong@ed.ac.uk}
\emailAdd{stefan.prohazka@ed.ac.uk}
\emailAdd{jsalzer@fas.harvard.edu}

\abstract{
We construct various limits of JT gravity, including Newton--Cartan and Carrollian versions of dilaton gravity in two dimensions as well as a theory on the three-dimensional light cone. In the BF formulation our boundary conditions relate boundary connection with boundary scalar, yielding as boundary action the particle action on a group manifold or some Hamiltonian reduction thereof. After recovering in our formulation the Schwarzian for JT, we show that AdS--Carroll gravity yields a twisted warped boundary action. We comment on numerous applications and generalizations.
}

\preprint{TUW--20--05}

\begin{document} 
\maketitle
\flushbottom

\section{Introduction}
\label{sec:introduction}

Jackiw--Teitelboim (JT) gravity
\cite{Barbashov:1980bm,Dhoker:1982wmk,Teitelboim:1983ux,Jackiw:1984je}
features prominently in classical and quantum gravity as a convenient
toy model to elucidate conceptual problems while keeping the technical
ones at a bare minimum. Examples include implementing 't Hooft's brick
wall proposal \cite{Mann:1989gh}, Cardyology attempts
\cite{Cadoni:1998sg}, noncommutative geometry
\cite{Cacciatori:2002ib}, holographic renormalization and
thermodynamics \cite{Grumiller:2007ju}, the attractor mechanism
\cite{Sen:2008yk}, constant dilaton holography
\cite{Hartman:2008dq,Castro:2008ms}, the JT/SYK correspondence
\cite{Maldacena:2016upp} (for reviews see
\cite{Mertens:2018fds,Sarosi:2017ykf,Gu:2019jub}), relations to random matrix
models \cite{Cotler:2016fpe,Saad:2019lba}, $T\bar T$-deformations
\cite{Dubovsky:2017cnj,Cardy:2018sdv}, traversable wormholes
\cite{Maldacena:2018lmt}, holographic complexity \cite{Goto:2018iay},
constructions of the Hartle--Hawking wavefunction
\cite{Harlow:2018tqv,Kitaev:2018wpr} and implementations of the island
proposal to resolve the black information loss problem
\cite{Penington:2019npb,Almheiri:2019psf,Almheiri:2019hni,Almheiri:2019yqk,Almheiri:2019qdq,Penington:2019kki}.
See \cite{Grumiller:2002nm} for a review on further aspects of
two-dimensional (2d) dilaton gravity, including numerous
generalizations of JT gravity, like the
Callan--Giddings--Harvey--Strominger (CGHS)
model~\cite{Callan:1992rs}. None of these generalizations so far gave
up the assumption of (pseudo-)Riemannian metrics (or a corresponding
Cartan formulation).

For applications or toy models of non-relativistic holography it is of
interest to consider singular limits of JT gravity to, say, Carrollian
or Galilean spacetimes. The main purpose of our work is to show how
this is done and to discuss some aspects of these new models,
including boundary actions and boundary conditions. Among other
applications our construction allows to address questions such as ``Is
there a Newton--Cartan version of 2d dilaton gravity?''
or ``What is the Schwarzian analogue for the AdS--Carroll limit of JT gravity?''.

The urge to look for theories beyond the Riemann--Cartan setup is
partly motivated by the relation of Carrollian symmetries to null
surfaces, like horizons or null infinity in flat space, and partly by
applications of non-relativistic theories in condensed matter physics.
Eventually, some of our models may serve as gravity duals for examples
of non- or ultra-relativistic holography in the spirit of the JT/SYK
correspondence, and many of the questions addressed and issues raised
in this context could potentially be transposed to models introduced
in our work.

In several ways this work mirrors investigations of Chern--Simons
theories in $2+1$ dimensions based on Lie algebras beyond the
semi-simple case, started in \cite{Achucarro:1987vz,Witten:1988hc} for
Poincar\'e and (A)dS and extended to
Galilei~\cite{Papageorgiou:2009zc,Papageorgiou:2010ud} and
beyond~\cite{Hartong:2016yrf,Bergshoeff:2016lwr,Hartong:2017bwq,Matulich:2019cdo}
and to higher spins~\cite{Bergshoeff:2016soe}.

This paper is organized as follows. In Section \ref{sec:choice-theory} we review general aspects of the formulation of 2d gravity as a gauge theory of BF-type. In Section \ref{sec:jt-gravity-as} we take singular limits of JT gravity, among other things to Galilean and Carrollian theories that we generalize to Newton--Cartan and Carroll dilaton gravity; we also show that there is a light cone theory that does not require any limit. In Section \ref{sec:limit} we focus on the subclass of metric BF theories and their limits. In Section \ref{sec:boundary} we discuss boundary actions and how to perform a Hamiltonian reduction from the action for a particle moving on a group manifold to a Schwarzian-like action by imposing certain constraints. In Section \ref{sec:schwarzianBF} we discuss two examples for such boundary actions, first for JT and then for AdS--Carroll$_2$ gravity. In Section \ref{sec:conclusions} we conclude with a discussion of possible applications and generalizations.

\subsection*{Notation}

When applicable, upper (lower) signs in equations refer to the Euclidean (Lorentzian) case.

\paragraph*{Note added:}

Shortly after our work~\cite{Gomis:2020wxp} appeared on the arXiv.
Where applicable, our results agree with each other.

\section{2d gauge theories of gravity}
\label{sec:choice-theory}

JT gravity in its first order formulation \cite{Isler:1989hq,Chamseddine:1989yz} is a specific BF theory based on the Lie algebra $\mathfrak{so}(2,1)$, which features an invariant metric. Some of its limits may lead to Lie algebras without metric.  Since these subtleties will be relevant for the remaining work, we set the stage by providing a rather detailed reminder of BF theories.

We follow~\cite{Salzer:2018zlv} where further details are
provided (see also the review~\cite{Birmingham:1991ty}; especially
relevant is Section 6 on Schwarz type topological gauge theories).

\subsection{BF theories}
\label{sec:choice-trace}

\textbf{BF theory} is defined by the bulk action
\begin{subequations}
  \label{eq:BF}
\begin{align}
  I^{\mathrm{BF}}[\mathcal{X}^{*},A]&=\frac{k}{2\pi}\int_{{\cal M}^2}\,\lagr^{\mathrm{BF}}[\mathcal{X}^{*},A] \\
   \lagr^{\mathrm{BF}}[\mathcal{X}^{*},A] &= \mathcal{X}^{*} F 
  = X_{K} 
  \Big(
  \dd A^{K} + \frac{1}{2} c\indices{_{IJ}^{K}} A^{I} \wedge A^{J} 
  \Big)  
\end{align}
\end{subequations}
where $k$ is a dimensionless coupling constant,
$\mathcal{X}^{*} = X_{I} \Et^{I}$ is a scalar transforming in the
coadjoint representation and the Lie algebra valued one-form
$A=A^{I}_{\mu} \ett_{I} \dd x^{\mu}$ is a gauge field with curvature
two-form $F \equiv \dd A + \frac{1}{2}[A,A]$. The structure constants
$c\indices{_{IJ}^{K}}$ of a Lie algebra $\mathfrak{g}$ are defined by
$[\ett_{I},\ett_{J}] = c\indices{_{IJ}^{K}}\ett_{K}$ (with the dual
$\mathfrak{g}^{*}$ with basis $\Et^{I}$ given by
$\Et^{I}(\ett_{J})=\delta^{I}_{J}$). For $X,Y \in \mathfrak{g}$ and
$\mathcal{Y} \in \mathfrak{g}^{*}$ the adjoint and coadjoint actions
are given by $\mathrm{ad}_{X}Y=[X,Y]$ and
$\mathrm{ad}_{X}^{*}\mathcal{Y}(\cdot) =- \mathcal{Y}( \mathrm{ad}_{X}
\cdot)$, respectively. Equivalently, in a basis this reads
$\mathrm{ad}_{\ett_{I}} \ett_{J}
=[\ett_{I},\ett_{J}]=c\indices{_{IJ}^{K}} \ett_{K}$ and
$\mathrm{ad}^{*}_{e_{I}}\Et^{J}= - c\indices{_{IK}^{J}}\Et^{K}$.
  
The definition of BF theory does not require a trace or invariant
metric. This is different from gauge theories of, e.g., Chern--Simons
or Yang--Mills type that are based on Lie algebras with an invariant
metric. The gauge transformations are given by
$\vd_{\lambda}\mathcal{X}^{*}=\mathrm{ad}^{*}_{\lambda}\mathcal{X}^{*}$
and $\vd_{\lambda}A = - (\dd \lambda + [A,\lambda])$ (and hence
$\vd_{\lambda}F = \mathrm{ad}_{\lambda}F = [\lambda,F]$), or
explicitly,
\begin{align}\label{eq:gaugBFind}
\vd_{\lambda} X_{I} &=  c\indices{_{IJ}^{K}} \lambda^{J}X_{K}  &
\vd_{\lambda}A^{I} &= - \dd \lambda^{I} - c\indices{_{JK}^{I}} A^{J} \lambda^{K} 
\end{align}
and leave invariant the action \eqref{eq:BF}. Varying it leads to the
equations of motion
\begin{align}
F^{I} = \dd A^{I} + \frac{1}{2} c\indices{_{JK}^{I}} A^{J} \wedge A^{K} &= 0  &
\dd X_{I} + c\indices{_{IJ}^{K}} A^{J} X_{K} = 0 \, .
\end{align}

An interesting subclass of BF theories is obtained when the gauge
algebra is given by a metric or (regular) quadratic Lie algebra. This
means that the Lie algebra admits an \emph{invariant metric} 
$\langle \cdot , \cdot \rangle: \mathfrak{g} \times \mathfrak{g} \to
\R$ that is a non-degenerate, symmetric, $\mathrm{ad}$-invariant
bilinear form (in the following we will often omit the comma; by
$\mathrm{ad}$-invariance we mean
$\langle [z,x],y\rangle+\langle x, [z,y]\rangle=0$ for all Lie algebra
elements $x,y,z \in \mathfrak{g}$). Since it is non-degenerate, we can
use this metric to identify elements of the dual $\mathfrak{g}^{*}$
with elements of the Lie algebra $\mathfrak{g}$ via
$\langle \mathcal{X} , \cdot \rangle = \mathcal{X}^{*}(\cdot)$, or
more explicitly $X^{I} = g^{IJ} X_{J}$ (where
$g_{IJ} = \langle \ett_{I}, \ett_{J}\rangle$ and
$g^{IJ}g_{JK}=\delta^{I}_{K}$). 

The Lagrangian for \textbf{metric BF theory} is given by
\begin{align}
  \label{eq:mBF}
  \lagr^{\mathrm{mBF}}[\mathcal{X},A] = \langle \mathcal{X}, F \rangle
    = g_{LK}X^{L}
  \Big(
  \dd A^{K} + \frac{1}{2} c\indices{_{IJ}^{K}} A^{I} \wedge A^{J}
  \Big)   
\end{align}
with equations of motion
\begin{align}
  \label{eq:EOM}
  F=0 \qquad\qquad \dd \mathcal{X} + [A,\mathcal{X}]=0\,.
\end{align}
As the coadjoint and adjoint representations are isomorphic we rewrite
the transformation
$\vd_{\lambda}\mathcal{X} =\mathrm{ad}_{\lambda}\mathcal{X}
=[\lambda,\mathcal{X}]$. The standard example of metric BF theories
are given by simple Lie algebras where one can use the matrix trace to
write $\lagr^{\text{mBF}} = \tr (\mathcal{X} F)$. There indeed exist
Lie algebras beyond the semisimple ones that admit an invariant
metric, e.g., one notable example is the $2+1$ dimensional Poincar\'e
algebra. There exists a structure theorem that helps to understand
them~\cite{Medina1985} (see also \cite{FigueroaO'Farrill:1995cy})
which has also been used to find generalizations for the Galilei and
Carrollian cases which we will encounter
below~\cite{Matulich:2019cdo}. We provide an overview of all Lie
algebras of low dimension that admit an invariant metric in
Appendix~\ref{sec:metric-lie}.

\subsection{Geometric interpretation}
\label{sec:choice-spacetime}

Up until now the BF theory \eqref{eq:BF}, or metric BF theory
\eqref{eq:mBF}, defines a (topological) gauge theory based on a gauge
algebra $\mathfrak{g}$ without any geometrical meaning. In this
subsection we clarify how the fields appearing in these actions
acquire a geometric interpretation upon introducing additional
structure in the form of a \emph{Klein pair}. To this end it is
convenient to introduce the notion of \emph{kinematical spacetime}.

As is well-known, (A)dS and Minkowski space, and quotients thereof,
are the only Lorentzian manifolds that are both homogeneous and
isotropic. In $n$ spacetime dimensions this implies the existence of
$n(n+1)/2$ Killing vectors. From this follows that one can describe
these spacetimes, without introducing a metric, as the homogeneous
spaces $G/SO(n-1,1)$ with $G=SO(n-1,2)$ for AdS, $G=SO(n,1)$ for dS,
and $G=ISO(n-1,1)$ for Minkowski, respectively. In addition to being
homogeneous, all these spacetimes are also isotropic. By this we mean
that the symmetry group contains $SO(n-1)$ as a subgroup and splits
into representations thereof.

More generally, one can ask the question which other homogeneous and
isotropic spaces, i.e., \emph{kinematical spacetimes}, of dimensions
$n$ exist. This question, studied first in \cite{Bacry:1968zf} and
answered exhaustively in \cite{Figueroa-OFarrill:2018ilb}, boils down
to a classification of so-called \emph{Klein pairs}
$(\mathfrak{g},\mathfrak{h})$ with $\mathfrak{h}$ an $n(n-1)/2$
dimensional subalgebra of the $n(n+1)/2$ dimensional Lie algebra
$\mathfrak{g}$ that together determine the respective kinematical
spacetimes. Note that most of these spacetimes do not exhibit a metric
of Lorentzian or Euclidean signature but can be classified according
to the existence of an ultra-relativistic \emph{Carrollian} or a
non-relativistic \emph{Galilean} structure. The former consists of a
degenerate metric whose kernel is spanned by a single vector field. A
Galilean structure, on the other hand, is defined by a degenerate
co-metric whose kernel is spanned by a nowhere vanishing one-form. One
can furthermore define, in most cases, a distinguished connection on
these spacetimes and classify these spacetimes according to the
curvature of this connection\footnote{For non-homogeneous Carrollian
  or Galilean spacetimes there does not exists a preferred connection,
  like the Levi-Civita connection in the case of Lorentzian geometries
  \cite{Hartong:2015zia,Bekaert:2015xua,Hartong:2015xda}.}. For more details we refer
the reader to \cite{Figueroa-OFarrill:2018ilb}. We emphasize again
that the choice of subalgebra $\mathfrak{h}$ is important as
spacetimes with isomorphic symmetry algebra $\mathfrak{g}$ can have
vastly different geometric properties depending on the choice of
$\mathfrak{h}$. We will encounter examples illustrating this fact
throughout this work.

Returning to the problem at hand, we denote the three generators of
$\mathfrak{g}$ in $1+1$ dimensions as $\mathfrak{g}=\{\Ht,\Pt,\Bt\}$
and use the convention that $\mathfrak{h}$ is \emph{always} spanned by
the generator denoted by $\Bt$. Expanding the Lie algebra-valued one-form of a BF theory based on this algebra as
\begin{equation}
  \label{eq:29}
  A= \tau \Ht+ e \Pt+ \omega \Bt
\end{equation}
we can now interpret the field associated to $\Ht$ ($\Pt)$ as temporal
(spatial) zweibein component. The field $\omega$ is the gauge field of
the internal symmetry transformation, i.e., the dualized
spin-connection associated to local Lorentz transformations in the
relativistic case. The Klein pair thus provides a map from the a
priori abstract gauge field to geometric data.

\section{Limits of JT gravity}
\label{sec:jt-gravity-as}

To set the stage we briefly review the well-known (A)dS JT gravity
case in BF formulation. Next we define a novel BF theory on the light
cone, which is based on the same simple Lie algebra, but the
underlying spacetimes differ due to different geometric
interpretations of the gauge connection components. Then we discuss
and provide the kinematical limits of (A)dS JT gravity.

\subsection{AdS and dS BF theory}
\label{sec:ads-vs.-ds}

The AdS ($\lc<0$) and dS ($\lc>0$) BF theories can be written in a
covariant fashion where we use $\Pt_{a}=(\Pt_{0},\Pt_{1})=(\Ht,\Pt)$.
They are based on the Lie algebra
  \begin{align}
  \label{eq:AdScov}
  [\Bt,\Pt_{a}] &=-\epsilon\indices{_{a}^{b}} \Pt_{b} &  [\Pt_{a},\Pt_{b}] &= - \lc \epsilon_{ab} \Bt
\end{align}
where $\epsilon_{01}=1$ and we raise and lower with
$\eta_{ab}=\mathrm{diag}(1,\,-1)$ in the Lorentzian and
$\eta_{ab}=\delta_{ab}=\mathrm{diag}(1,\,1)$ in the Euclidean case.
The most general invariant metric is
\begin{align}
  \label{eq:adsmetrfixed}
 \langle \Bt,\Bt \rangle&= \mu   &
 \langle \Pt_{a},\Pt_{b} \rangle&=  \mu \lc\eta_{ab} 
\end{align}
where $\mu\neq 0$ is an overall proportionality factor.

The Lie algebras for positive and negative $\lc$ are isomorphic. It is
the choice which generator is part of the spin-connection and which is
part of the vielbein that provides the distinction between these two
cases. As discussed in the previous section, assuming that $\Bt$ spans
the subalgebra $\mathfrak{h}$, the Klein pair dictates the form
\begin{align}
  \label{eq:fieldsSL2}
  A &= \omega \Bt + e^{a} \Pt_{a} &
  \mathcal{X} &= X \Bt + X^{a}\Pt_{a}
\end{align}
of the gauge field and the coadjoint scalar. Together with the metric
\eqref{eq:adsmetrfixed} this allows us to write the Lagrangian for
(A)dS-JT gravity as
\begin{equation}
  \label{eq:JTlagrangian}
\mathcal{L}_{\mathrm{JT}}=X\Big(\dd \omega-\frac{\lc}{2}e^{a}\wedge e^{b}\epsilon_{ab}\Big)+ X^a\Big(\dd e_a+\epsilon\indices{_a^b}\omega \wedge e_b\Big)
\end{equation}
where we rescaled $\lc X^a\to X^a$ for convenience. The fields $X^a$
enforce the 2d torsion constraint for the zweibein $e^a$.
Upon solving this for the spin connection $\omega$ and plugging it
into the action, one is left with the well-known second-order action
for JT gravity with $X$ being the dilaton field.

As a reminder the Penrose diagrams of (A)dS$_2$ are depicted in
Fig.~\ref{fig:2}, where the transformations generated by translations
$\Ht$, $\Pt$ and boosts $\Bt$ are indicated.

\def \L {1.4}
\begin{figure}[htb]
    \centering
    \begin{tikzpicture}
     \draw[thick,black] (-2*\L,-1*\L) -- (-2*\L,1*\L);
     \draw[thick,black] (-1*\L,-1*\L) -- (-1*\L,1*\L);
     \draw[black] (-1.1*\L,0) node[left] (scrip) {AdS$_2$};
     \draw[red,->] (-0.85*\L,-0.25*\L) -- (-0.85*\L,0.25*\L);
     \draw[red] (-0.85*\L,0*\L) node[right] (scrip) {$\Ht$};
     \draw[blue,->] (-1.75*\L,-1*\L) -- (-1.25*\L,-1*\L);
     \draw[blue] (-1.5*\L,-1*\L) node[above] (scrip) {$\Pt$};
     \draw[black,dotted] (-2*\L,0.75*\L) -- (-1*\L,0.75*\L);
     \draw[black,dotted] (-2*\L,0.55*\L) -- (-1*\L,0.95*\L);
     \draw[thick,black,dotted,->] (-2.15*\L,0.8*\L) -- (-2.15*\L,0.5*\L);
     \draw[black] (-2.15*\L,0.65*\L) node[left] (scrip) {$\Bt$};
     \draw[thick,black] (0.5*\L,-0.5*\L) -- (2.5*\L,-0.5*\L);
     \draw[thick,black] (0.5*\L,0.5*\L) -- (2.5*\L,0.5*\L);
     \draw[black] (1.8*\L,0) node[left] (scrip) {dS$_2$};
     \draw[red,->] (2.65*\L,-0.25*\L) -- (2.65*\L,0.25*\L);
     \draw[red] (2.65*\L,0*\L) node[right] (scrip) {$\Ht$};
     \draw[blue,->] (0.65*\L,-0.4*\L) -- (1.15*\L,-0.4*\L);
     \draw[blue] (0.9*\L,-0.4*\L) node[above] (scrip) {$\Pt$};
     \draw[black,dotted] (0.5*\L,0.2*\L) -- (2.5*\L,0.8*\L);
     \draw[black,dotted] (0.5*\L,-0.8*\L) -- (2.5*\L,-0.2*\L);
     \draw[thick,black,dotted,->] (0.35*\L,0.55*\L) -- (0.35*\L,0.15*\L);
     \draw[black] (0.35*\L,0.35*\L) node[left] (scrip) {$\Bt$};
    \end{tikzpicture}
    \caption{Penrose diagrams for AdS$_2$ (left) and dS$_2$ (right).}
    \label{fig:2}
\end{figure}
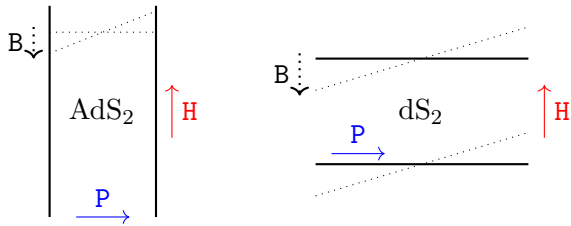

For later purposes we discuss now briefly how to arrive at arbitrary
dilaton gravity models starting from JT. The JT Lagrangian
\eqref{eq:JTlagrangian} has as most general Lorentz invariant
generalization that preserves the Palatini condition of vanishing
on-shell torsion the Lagrangian
\begin{equation}
\mathcal{L}_{\mathrm{dil}}=X\left(\dd \omega+V(X)\,e^{a}\wedge e^{b}\epsilon_{ab}\right)+ X^a\left(\dd e_a+\epsilon\indices{_a^b}\omega \wedge e_b\right)\,.
\label{eq:jt1}
\end{equation}
that depends on an arbitrary function of the dilaton field, $V(X)$.
(Dropping the Palatini condition further generalizes
$V(X)\to V(X,\,X^a X_a)$.) Thus, JT naturally generalizes to generic
dilaton gravity models. The same is true for its various limits
studied below. 

\subsection{BF on the light cone}
\label{sec:bf-light-cone}

In addition to the well-known homogeneous spaces and their BF theories
mentioned in the previous subsection there exists another homogeneous
space based on the symmetry algebra $\mathfrak{sl}(2,\,\mathbb{R})$
which is the light cone of three dimensional Minkowski space seen as
2d manifold. The homogeneous space of the light cone is
based on the following algebra, see,
e.g.,~\cite{Figueroa-OFarrill:2018ilb}
  \begin{align}
  [\Bt,\Ht] &= -\Bt &
  [\Bt,\Pt] &= \Ht &
  [\Ht,\Pt] &= - \Pt
  \label{eq:LC}
  \end{align}
and invariant metric
\begin{align}
  \label{eq:LCmetr}
 \langle \Bt,\Pt \rangle&=\mu & \langle \Ht,\Ht \rangle&=  \mu \, .
\end{align}
The Lie algebra is, for any dimension, isomorphic to the one of de
Sitter spacetime~\cite{Figueroa-OFarrill:2018ilb} (in $1+1$ dimensions
also to the one of anti-de Sitter). However due to the different
choice of subalgebra $\mathfrak{h}$, the corresponding homogeneous
spaces differ. In particular, the light cone is a Carrollian
spacetime, i.e., one can define an invariant Carrollian structure on this spacetime; see \cite{Figueroa-OFarrill:2018ilb} for the explicit construction.

In a sense the light cone is in between AdS$_2$ (which has timelike
asymptotic boundaries) and dS$_2$ (which has spacelike asymptotic
boundaries). Instead of a Penrose diagram we just depict the light cone
itself in Fig.~\ref{fig:3}, together with the geometric interpretation
of the three generators.

\def \L {1.4}
\def \elL {2.025*\L}
\def \elW {0.45*\L}
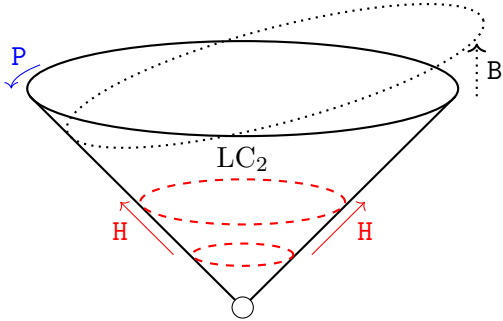
\begin{figure}[htb]
    \centering
    \begin{tikzpicture}
     \draw[thick,black] (-0.07*\L,-0.93*\L) -- (-2*\L,1*\L);
     \draw[thick,black] (0.07*\L,-0.93*\L) -- (2*\L,1*\L);
     \draw[black] (0,-1*\L) circle (0.1*\L);
     \draw[thick,black] (0,1.073*\L) ellipse ({\elL} and {\elW});
     \draw[black] (0,0.62*\L) node[below] (scrip) {LC$_2$};
     \draw[red,thick,dashed] (0,-0.5*\L) ellipse ({0.235*\elL} and {0.235*\elW});
     \draw[red,thick,dashed] (0,0*\L) ellipse ({0.475*\elL} and {0.475*\elW});
     \draw[red,->] (0.65*\L,-0.5*\L) -- (1.15*\L,0*\L);
     \draw[red] (1.15*\L,-0.1*\L) node[below] (scrip) {$\Ht$};
     \draw[red,->] (-0.65*\L,-0.5*\L) -- (-1.15*\L,0*\L);
     \draw[red] (-1.15*\L,-0.1*\L) node[below] (scrip) {$\Ht$};
     \draw[blue,->] (-0.15*\L,1.073*\L) [partial ellipse=150:180:{\elL} and {\elW}];
     \draw[blue] (-2.1*\L,1.2*\L) node[above] (scrip) {$\Pt$};
     \draw[black,thick,dotted,rotate=15] (0.605*\L,1.073*\L) ellipse ({\elL} and {\elW});
     \draw[black,thick,dotted,->] (2.2*\L,1.0*\L) -- (2.2*\L,1.5*\L);
     \draw[black] (2.2*\L,1.25*\L) node[right] (scrip) {$\Bt$};
    \end{tikzpicture}
    \caption{Two-dimensional future light cone of three-dimensional
      Minkowski space with vertex removed. Topologically, this is a
      strip like (A)dS$_2$.}
    \label{fig:3}
\end{figure}

The light cone theory has the Lagrangian
\begin{equation}
  {\mathcal L}_{\textrm{\tiny LC}}
  =\langle \mathcal{X},\,F\rangle
  = X^\Ht\,\big(\dd\tau+\omega\wedge e\big) + X^\Pt\,\big(\dd e-\tau\wedge e\big) +  X\,\big(\dd\omega-\omega\wedge\tau\big)
    \label{eq:lc1}
\end{equation}
where we used
\begin{equation}
\mathcal{X} = X^\Pt\,\Bt+X^\Ht\,\Ht+X\,\Pt\qquad\qquad A=\omega\,\Bt+\tau\Ht+e\Pt\,.
    \label{eq:lc2}
\end{equation}
This Lagrangian is equivalent to the JT Lagrangian, but the interpretation is different since `boosts' generated by $\Bt$ act differently in these two theories (see Figs.~\ref{fig:2} and \ref{fig:3}).

\subsection{Kinematical limits of BF theories}
\label{sec:limits-bf-theories}

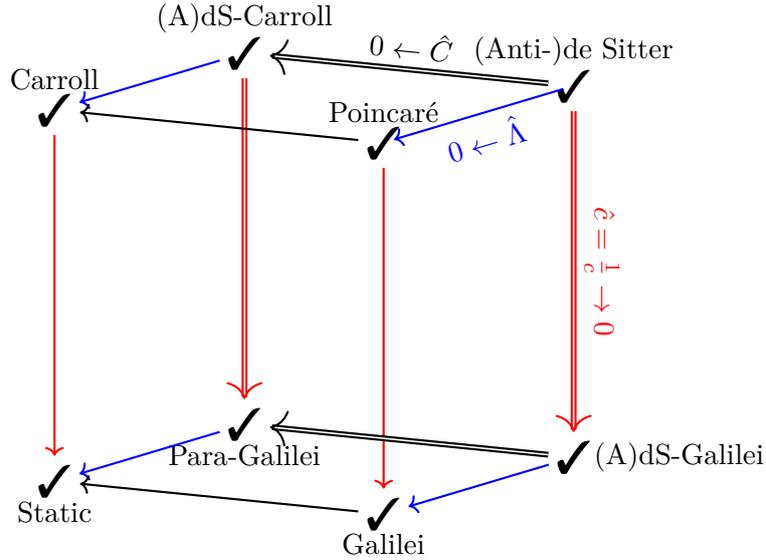
\begin{figure}[hbt]
  \centering
\tdplotsetmaincoords{80}{120}
\begin{tikzpicture}[
tdplot_main_coords,
dot/.style={circle,fill},
linf/.style={thick,->,blue},
cinf/.style={thick,->,red},
tinf/.style={thick,->},
stinf/.style={thick,->,gray},
shrink/.style={thick, ->,gray},
scale=0.25
]
\node (adsu1) at (-\shi ,10+\shi,10+\shi) [label=center:{\cmark}, label=above:(Anti-)de Sitter] {};
\node (pu1) at (10+\shi,10+\shi,10+\shi) [label=center:{\cmark}, label=above:Poincar\'e] {};
\node (nhu1label) at (-\shi,10+\shi,-\shi)  [label=center:{\cmark},label={right: (A)dS-Galilei}] {};
\node (nhu1) at (-\shi,10+\shi,-\shi) [label=center:{\cmark}] {};
\node (ppu1) at (-\shi,-\shi,10+\shi) [label=center:{\cmark}, label=above:(A)dS-Carroll] {};
\node (gu1) at (10+\shi,10+\shi,-\shi) [label=center:{\cmark}, label=below:Galilei] {};
\node (pgu1) at (-\shi,-\shi,-\shi) [label=center:{\cmark}, label=below:Para-Galilei] {};
\node (caru1) at (10+\shi,-\shi,10+\shi) [label=center:{\cmark}, label=above:Carroll] {};
\node (stu1) at (10+\shi,-\shi,-\shi) [label=center:{\cmark}, label=below:Static] {};

\draw[linf] (adsu1) -- node [sloped,below]{$0 \leftarrow \lc$}  (pu1);

\draw[shorten >=0.2cm,shorten <=0.2cm,linf] (nhu1) -- (gu1);
\draw[shorten >=0.2cm,shorten <=0.2cm,linf] (ppu1) -- (caru1);
\draw[shorten >=0.2cm,shorten <=0.2cm,linf] (pgu1) -- (stu1);

\draw[shorten >=0.2cm,shorten <=0.2cm,cinf,double] (adsu1) -- node [sloped,above] {$\cc=\frac{1}{c} \to 0$} (nhu1);
\draw[shorten >=0.2cm,shorten <=0.2cm,cinf,double] (ppu1) -- (pgu1);
\draw[shorten >=0.2cm,shorten <=0.2cm,cinf] (pu1) -- (gu1);
\draw[shorten >=0.2cm,shorten <=0.2cm,cinf] (caru1) -- (stu1);

\draw[shorten >=0.2cm,shorten <=0.2cm,tinf,double] (adsu1) -- node [sloped,above] {$0 \leftarrow \tc$} (ppu1);
\draw[shorten >=0.2cm,shorten <=0.2cm,tinf] (pu1) -- (caru1);
\draw[shorten >=0.2cm,shorten <=0.2cm,tinf] (gu1) -- (stu1);
\draw[shorten >=0.2cm,shorten <=0.2cm,tinf,double] (nhu1) -- (pgu1);

\end{tikzpicture}  
\caption{Kinematical limits of (anti-)de Sitter:
  non-relativistic/Galilean ($\cc=\frac{1}{c} \to 0$); flat
  ($\lc \to 0$); ultra-relativistic/Carrollian ($\tc \to 0$). Arrows
  with two lines imply there are two different limits, depending on
  the sign of the cosmological constant. }
\label{fig:cube}
\end{figure}

As discussed in Section \ref{sec:choice-trace} BF theories allow for a
gauge invariant action, irrespective of the existence of an invariant
metric on the Lie algebra. This means we can take limits without
compromising the well-definedness of the action and theory. Before we
show this we introduce an additional generator $\Mt$ into our theory
such that the nonzero commutators are given by
\begin{align}
  \label{eq:BFstlim}
  [\Bt,\Ht] &= \mp \tc^{2}  \Pt &
  [\Bt,\Pt] &= \cc^{2} \Ht + \alpha \Mt &
  [\Ht,\Pt] &= - \lc  \Bt  \, .
\end{align}
The upper (lower) sign specifies that the theory is Euclidean
(Lorentzian) in case this distinction is applicable and $\lc$ is the
cosmological constant which is (negative) positive for (A)dS
spacetimes. Each of the flat ($\lc \to 0$), Galilean
($\cc =\frac{1}{c}\to 0$)\footnote{The inverse speed of light $\cc$ is
  introduced so that all contractions involve parameters tending to
  zero.} or Carrollian ($\tc \to 0$) limits leads again to a
well-defined Lie algebra; for a visualization of these limits see
Fig.~\ref{fig:cube}. To reduce clutter, the central extension $\Mt$ was dropped
in this diagram and the Euclidean cases and the light cone algebra of the previous section, that does not follow from any limit, are not represented in this diagram.

The new generator $\Mt$ for $\cc \neq 0$ is a trivial central
extension that before taking any limits could be eliminated by a shift
of $\Ht$, which shows that the starting point is actually the direct
sum $\mathfrak{sl}(2,\R) \oplus \mathfrak{u}(1)$. However, it becomes
a nontrivial central extension in the Galilean limit
$\cc \to 0$. We refer to the centrally extended Galilean algebra with
nonzero $\lc$ as the extended (A)dS-Galilei algebra (also sometimes
referred to as the Newton--Hooke algebra). Sending $\lc \to 0$ leads
to the centrally extended Galilean algebra, better known as the
Bargmann algebra. We could have added a similar central extension on
the right hand side of each of the other brackets leading to centrally
extended Carrollian and Poincar\'e algebras.

The contracted action, equations of motion and gauge symmetries are
well defined as long as the Lie algebra contraction is well defined.
We have summarized further possibly interesting algebras that do not
follow from a kinematical limit, like Lifshitz, Schr\"odinger and
$1/c$ expanded Poincar\'e in Appendix~\ref{sec:lifshitz-schrodinger}.

We now construct the limits of the action, equations of motion, and
gauge transformations explicitly and study their Lorentzian, Galilean
and Carrollian invariants. The coadjoint scalar $\mathcal{X}^{*}$ and
the gauge connection $A$ are parametrized as
\begin{align}
  \mathcal{X}^{*} &= X \Bt^{*} + X_{\Ht} \Ht^{*} + X_{\Pt} \Pt^{*} + X_{\Mt} \Mt^{*}
  &   A &= \omega \Bt + e^{a} \Pt_{a} + m \Mt  = \omega \Bt + \tau \Ht + e \Pt + m \Mt 
\end{align}
where the dual basis is defined by
\begin{align}
  \Bt^{*}(\Bt) =1 \qquad \Ht^{*}(\Ht) = 1 \qquad \Pt^{*}(\Pt) =1 \qquad \Mt^{*}(\Mt) =1 \, .
\end{align}
The Lagrangian is given by
\begin{align}\label{eq:Lagrangian:so(1,2) plus u(1)}
  \lagr[\mathcal{X}^{*}, A] = X F(\Bt)  + X_{\Ht}F(\Ht) +X_{\Pt} F(\Pt)  + X_{\Mt} F(\Mt)
\end{align}
with curvature
\begin{subequations}
\label{eq:coadF=0}
\begin{align}
  F& = F(\Bt) \Bt + F(\Ht) \Ht + F(\Pt) \Pt + F(\Mt)\Mt \\
     &= (\dd \omega - \lc \tau \wedge e)\Bt
     + (\dd \tau + \cc^{2} \omega \wedge e) \Ht
       + (\dd e \mp \tc^{2} \omega \wedge \tau) \Pt
       + (\dd m + \alpha \omega \wedge e)\Mt \, .
\end{align}
\end{subequations}
The action is invariant under the gauge transformations parametrized
by
$\lambda = \lambda^{\Bt} \Bt + \lambda^{\Ht} \Ht + \lambda^{\Pt}\Pt +
\lambda^{\Mt}\Mt$,
\begin{subequations}
\label{eq:coadsym}
\begin{align}
  \vd_{\lambda} A & =
  \left[
  - \dd \lambda^{\Bt} + \lc (\tau \lambda^{\Pt} - e \lambda^{\Ht})
                                                  \right] \Bt
   +
  \left[
  - \dd \lambda^{\Ht} - \cc^{2} (\omega \lambda^{\Pt} - e \lambda^{\Bt} )
  \right] \Ht
  \nonumber \\
 & \quad + 
  \left[
  - \dd \lambda^{\Pt} \pm \tc^{2} (\omega \lambda^{\Ht} - \tau \lambda^{\Bt})
  \right] \Pt
  +
  \left[
  - \dd \lambda^{\Mt} - \alpha (\omega \lambda^{\Pt} - e \lambda^{\Bt} )
   \right] \Mt
   \\
  \vd_{\lambda} \mathcal{X}^{*} 
  &=\left[
  \mp \tc^{2} X_{\Pt} \lambda^{\Ht} +(\alpha X_{\Mt} + \cc^{2} X_{\Ht}) \lambda^{\Pt}
  \right] \Bt^{*}
  +
  \left[
   \pm \tc^{2} X_{\Pt} \lambda^{\Bt} - \lc X \lambda^{\Pt}
  \right] \Ht^{*} \nonumber \\
  &\quad +
  \left[
   \lc X \lambda^{\Ht} - (\alpha X_{\Mt} + \cc^{2} X_{\Ht}) \lambda^{\Bt}
  \right] \Pt^{*} \, .
\end{align}
\end{subequations}
All contraction parameters have positive exponent and consequently the limits are well-defined. The equations of motion of \eqref{eq:Lagrangian:so(1,2) plus u(1)} are $F=0$ and
\begin{align}
\label{eq:coaddX=}
  \dd\mathcal{X}^{*}+ \mathrm{ad}_{A}^{*}\mathcal{X}^{*}
  &=\left[
  \dd X \mp \tc^{2} X_{\Pt} \tau + (\alpha X_{\Mt} + \cc^{2} X_{\Ht}) e
  \right] \Bt^{*}
  +
  \left[
  \dd X_{\Ht} \pm \tc^{2} X_{\Pt} \omega -\lc X e
  \right] \Ht^{*} \nonumber \\
  &\quad +
  \left[
  \dd X_{\Pt} + \lc X \tau - (\alpha X_{\Mt} + \cc^{2} X_{\Ht}) \omega
    \right] \Pt^{*}
    + \dd X_{\Mt} \, \Mt^{*}  = 0\,.
\end{align}

Since they provide us with additional geometric information, we first investigate the invariants of the local boosts
\begin{align}
  \vd_{\lambda^{\Bt}} \tau &=\cc^{2} \lambda^{\Bt} e  &
  \vd_{\lambda^{\Bt}} e &= \mp  \tc^{2} \lambda^{\Bt} \tau \, .
\end{align}
Expectedly, these transformations are independent of curvature, as evident from their independence of the parameter $\lc$. Without taking any limit neither $\tau$ nor $e$ is invariant, which is familiar from Lorentzian geometry where no distinguished invariant vector field or one-form exists. In the Galilean limit $\cc \to 0$ we get the invariant `clock one-form' $\tau$. In the Carrollian limit the spatial zweibein component $e$ is invariant. Only in the Lorentzian/Euclidean case we can define the invariant Lorentzian/Euclidean \emph{non-degenerate} metric
\begin{align}
  g = \mp  \tau^{2} + e^{2} \,.
\end{align}
This is a manifestation of the fact that we are looking beyond
Lorentzian geometries, and it justifies our claim to define Galilean
and Carrollian gravitational theories. Additionally this shows that
theories based on the same algebra can be geometrically different, but
are connected via dualities upon exchanging, e.g., time and space as
in the case of Galilei and Carroll. This should be viewed as a local
statement, since globally the spatial direction may be compact and the
time direction non-compact.

We will study the equations of motion of the Lagrangian
\eqref{eq:Lagrangian:so(1,2) plus u(1)} for the special cases
$\hat c=0$ or $\hat C=0$. When $\hat c$ and $\hat C$ are both nonzero
the equations setting the $\Ht$ and $\Pt$ curvatures equal to zero are
the zero torsion conditions in the Cartan description of
Euclidean/Lorentzian geometries. The zero torsion equations can be
solved for $\omega$. The abelian curvature of $\omega$, i.e.,
$d\omega$, is then either set equal to the volume form (for
$\hat\Lambda\neq 0$) or to zero (for $\hat\Lambda=0$), corresponding
to the usual 1+1 dimensional maximally symmetric Euclidean/Lorentzian
spaces. There is also a zero curvature abelian gauge field, namely
$m-\frac{\alpha}{\hat c^2}\tau$.
 
 When $\hat C=0$ we are dealing with the following curvature equations 
 \begin{subequations}
 \begin{align}
 \dd \tau + \cc^{2} \omega \wedge e & =  0\\
 \dd e & =  0\\
 \dd m + \alpha \omega \wedge e & =  0\\
  \dd \omega -  \lc \tau \wedge e  & = 0 
\end{align}
\end{subequations}
which correspond to a constant curvature 2-dimensional Carrollian geometry with a zero curvature $U(1)$ gauge field on it that is given by $m-\frac{\alpha}{\hat c^2}\tau$. The fact that $\dd e=0$ can be interpreted as vanishing extrinsic curvature. To see this, one evaluates the 2-form $\dd e$ on the two vectors that are dual to the 1-forms $\tau$ and $e$. This leads to an expression involving the Lie derivative of the Carrollian metric $ee$ along the vector $v$ that spans the kernel of $ee$ (see \cite{Hartong:2015usd,Jensen:2017tnb} for a general discussion of 2-dimensional Carrollian geometries). The vanishing of $\dd e=0$ can also be rephrased as saying that the intrinsic torsion is zero \cite{Figueroa-OFarrill:2020gpr}. Unlike in the Lorentzian case, for Carrollian geometries we cannot solve for $\omega$ in terms of the vielbeine (and possibly the $m$ connection), since the curvature equations do not fix the vielbein component of $\omega$ along $e$, which is thus an independent field. This is not uncommon for Carrollian geometries. For example in \cite{Bergshoeff:2017btm} the undetermined components of $\omega$ were shown to correspond to Lagrange multipliers enforcing the constraint that the extrinsic curvature vanishes. In our setting the vanishing of the extrinsic curvature 
results from varying $X_{\Pt}$ in the Lagrangian. The role of $\omega$ is to ensure that the curvatures are Carroll boost invariant.

We next consider the case $\hat c=0$, setting $\tc=\alpha=1$ without loss of generality. In this case the $F=0$ equations of motion read
\begin{subequations}
\label{eq:c=0eqs}
\begin{align}
   \dd \tau & =  0 \\
   \dd e \mp  \omega \wedge \tau &= 0 \\
   \dd m + \omega \wedge e &= 0 \\
   \dd \omega -  \lc \tau \wedge e &= 0\,.
\end{align}
\end{subequations}
We interpret these equations in the language of Newton--Cartan (NC)
geometry. The second and third equation are the curvature constraints
imposed to be able to fully solve for the boost connection $\omega$ in
terms of the NC fields $\tau$, $e$ and $m$ \cite{Andringa:2010it}. The
remaining two equations fix the NC geometry.  The top equation states that the clock one-form $\tau$ is closed and so
these Newton--Cartan spaces admit absolute time (provided there are no
closed time circles so that $\tau$ is exact). The last
equation fixes the boost curvature which can be viewed as the
$1+1$ dimensional version of the equation for NC gravity. It is
interesting to point out that in 2d we can formulate a
Lagrangian theory of NC gravity (coupled to scalars) whereas in $2+1$
dimensions the Chern--Simons formulation of NC gravity requires an
additional connection related to a generator that is not contained
within the Bargmann algebra
\cite{Papageorgiou:2009zc,Bergshoeff:2016lwr,Hartong:2016yrf}. The
$1+1$ dimensional case with nonzero $\lc$ is based on the extended
(A)dS Galilei algebra, also called (extended) Newton--Hooke algebra,
which in $1+1$ dimensions admits an invariant metric. The case with
$\lc=0$ leads to the Bargmann algebra which does not admit an
invariant metric.

\subsection{Newton--Cartan dilaton gravity and Carroll dilaton gravity}
\label{sec:NCandCdilaton}

The Lagrangian \eqref{eq:Lagrangian:so(1,2) plus u(1)} with all contraction parameters set to unity is identical to the Lagrangian of the JT model \eqref{eq:JTlagrangian}, after removing the $\mathfrak{u}(1)$ field $\Mt$ by a redefinition of the generator $\Ht$. As discussed above, the Lagrangian \eqref{eq:Lagrangian:so(1,2) plus u(1)} with $\hat{c}=0$ defines a NC geometry. It is then fair to ask what is the NC version of more generic dilaton gravity \eqref{eq:jt1} (or its torsionful generalization below that equation)? 

We propose that this generalization is given by the NC dilaton gravity Lagrangian
\begin{equation}
 \lagr_{\tiny \textrm{NCdil}} = X\, \dd \omega+ X_\Ht\, \dd \tau+ X_\Pt\,(\dd e\mp\omega \wedge \tau)+X_\Mt\,(\dd m + \omega \wedge e) +V(X,\,X_\Pt)\,\tau\wedge e\,.
\label{eq:NCdilnew}
\end{equation}
where $V(X,\,X_\Pt)\propto X$ for the NC limit of JT and an arbitrary function more generally. The $\mp$ signs in front of $\omega$ in the third term can be exchanged by a field redefinition, so they do \emph{not} denote a distinction between a Lorentzian or Euclidean version of NC dilaton gravity. We nevertheless keep them to show they do not matter.

Only two of the curvature equations \eqref{eq:c=0eqs} change, but we nevertheless display all for convenience on the left hand side below, together with the remaining half of the equations of motion on the right hand side.
\begin{subequations}
\label{eq:NCeom}
  \begin{align}
    \dd\tau & = 0 & \dd X_\Ht \pm \omega \,X_\Pt+e\,V(X,\,X_\Pt)&=0 \\ 
    \dd e \mp \omega \wedge \tau + \partial_{X_\Pt} V(X,\,X_\Pt)\,\tau\wedge e & = 0 & \dd X_\Pt-\omega \,X_\Mt-\tau\, V(X,\,X_\Pt) &= 0\\
    \dd m + \omega \wedge e & = 0 & \dd X_\Mt & =  0\\
    \dd \omega+\partial_X V(X,\,X_\Pt)\,\tau\wedge e & = 0  & \dd X \mp X_{\Pt}\, \tau + X_{\Mt}\, e & = 0
\end{align}
\end{subequations}
The first equation on the left implies that the NC clock 1-form locally still is given by $\tau=\extd t$. One of the two curvature constraints (the second equation on the left) is modified if the potential $V$ depends on $X_\Pt$, while the other one (the third equation on the left) is unchanged. The last equation on the left controls the curvature of the geometry. For functions $V$ that are non-linear in $X$ the curvature is not constant. The equations on the right hand side are the NC version of the dilaton equations of motion. As in ordinary dilaton gravity (see, e.g.,~\cite{Grumiller:2006rc}) there is a `constant dilaton' and a `linear dilaton' sector. 
\begin{align}
    &\textrm{constant\;dilaton}: & X_\Mt &=0=X_\Pt & X &= X^c & X_\Ht &= X_\Ht^c \\
    &\textrm{linear\;dilaton}: & X_\Mt &= {\rm const.} & X &= X(t,\,x) & X_{\Pt,\,\Ht} &= X_{\Pt,\,\Ht}(t,\,x)
\end{align}
The constant $X^c$ is determined as roots of the potential,
    $V(X^c,\,0) = 0$.
As in ordinary dilaton gravity, curvature is constant for constant dilaton solutions
\begin{equation} 
\ast\dd\omega = -\partial_XV(X,\,X_\Pt)\, .
    \label{eq:cdv1}
\end{equation}
We defined our orientation by $\ast(\tau\wedge e)=1$.

The Lagrangian \eqref{eq:NCdilnew} is invariant under a non-linear modification of the symmetries~\eqref{eq:coadsym}
\begin{subequations}
  \label{eq:nonlin}
\begin{align}
  \delta \tau & =  -\dd \lambda^\Ht  &  \delta X_\Ht &=  X_\Pt \lambda^\Bt+V \lambda^\Pt\\
  \delta e &=-\dd\lambda^\Pt\pm\omega\lambda^\Ht\mp\tau\lambda^\Bt -\partial_{X_\Pt}V(\tau\lambda^\Pt-e\lambda^\Ht)& \delta X_\Pt & =  -V \lambda^\Ht-X_\Mt \lambda^\Bt \\
  \delta m &= -\dd\lambda^\Mt-\omega\lambda^\Pt+e\lambda^\Bt & \delta X_\Mt &= 0 \\
  \delta \omega &=-\dd \lambda^\Bt - \partial_XV (\tau \lambda^\Pt-e \lambda^\Ht)  & 
  \delta X &= \mp X_\Pt\lambda^\Ht+X_\Mt\lambda^\Pt \, .
\end{align}
\end{subequations}
Choosing a field-dependent parametrization of the gauge transformations,  $\lambda=\xi \cdot A+ \lambda^\Bt \Bt+\lambda^\Mt \Mt$, where $\xi$ is a 2d vector field, establishes that gauge transformations reduce on-shell to diffeomorphisms generated by $\xi$, Galilean boosts $\lambda^\Bt$, and $\mathfrak{u}(1)$ transformations $\lambda^\Mt$.

The general solution of NC dilaton gravity in the linear dilaton sector is obtained as follows [for simplicity we restrict to the torsionless case $V=V(X)$]. First we solve the clock 1-form equation $\dd\tau=0$ by $\tau=\dd t$, which partly gauge fixes diffeomorphisms. Next, we gauge fix boost invariance by demanding $e=K(t,\,x)\,\dd x$. Most of the remaining diffeomorphism invariance is fixed by setting $K(t,\,x)=1$. Finally, there is another abelian gauge symmetry generated by $\lambda^\Mt$ that we exploit to fix the mass 1-form as $m=\Phi(t,\,x)\,\dd t$, where $\Phi$ is interpreted as Newton potential. At this stage the residual gauge transformations are trivial coordinate shifts, $t\to t+t_0$ and $x\to x+x_0$, and time-dependent shifts generated by $\mathfrak{u}(1)$ transformations with $\lambda_\Mt=f(t)$ and boosts accompanied by compensating spatial diffeomorphisms with $\lambda_\Bt=\mp\partial_t\xi^x(t)=g(t)$. The latter two can be used to fix to zero integration functions encountered below.

Our 1-forms read
\begin{equation}
\tau = \dd t\qquad e = \dd x \qquad m = \Phi(t,\,x)\,\dd t \qquad \omega =\partial_x\Phi(t,\,x)\,\dd t\,.
    \label{eq:angelinajolie}
\end{equation}
The result \eqref{eq:angelinajolie} solves all the equations of motion on the left side of \eqref{eq:NCeom} provided the Newton potential $\Phi$ obeys the second order partial differential equation
\begin{equation}
\partial_x^2\Phi(t,\,x) = \frac{\dd V(X)}{\dd X}
    \label{eq:lalapetz}
\end{equation}
If $V=\rm const.$ we are back to the NC case with cosmological constant and obtain the expected confining potential, $\Phi\propto x$, after setting to zero all integration functions by residual gauge fixing.

The right half of the equations of motion \eqref{eq:NCeom} is solved as follows. The penultimate one trivially yields $X_\Mt=-1/m_0=\rm const$. The other three equations combined yield the non-linear (Casimir) relation
\begin{equation}
X_\Mt X_\Ht \pm\frac12\,X_\Pt^2 = w(X)\qquad\qquad w(X):=\int^XV(y)\dd y\,.
\label{eq:NCC}
\end{equation}
The last equation of motion on the right \eqref{eq:NCeom} integrates to $X=x/m_0+x_0(t)$ where $\dot x_0(t)=\pm X_\Pt(t)$. We exploit now the residual gauge transformation generated by $\lambda_\Bt=\mp\partial_t\xi^x(t)=g(t)$ to gauge fix $X_\Pt=0$, yielding $\partial_X\Phi = V$ when plugged into the second equation of motion on the right \eqref{eq:NCeom}. Integrating this equation yields $\Phi=w(X)+\Phi_0(t)$, where the integration function $\Phi_0$ can be gauge fixed to zero with our remaining residual gauge transformation generated by $\lambda_\Mt=f(t)$. Finally, the relation \eqref{eq:NCC} can be solved for $X_\Ht$.

Our solution
\begin{equation}
X = \frac{x}{m_0}\qquad X_\Mt=-1\qquad X_\Pt=0\qquad X_\Ht = - m_0\,w(X)
    \label{eq:NCdilaton}
\end{equation}
shows that the label `linear dilaton sector' is indeed justified, as the dilaton is linear in the spatial coordinate $x$. It contains one relevant constant of motion, $m_0$, related to the mass of the solution. In the chosen gauge the Newton potential
\begin{equation}
    \Phi(X) = w(X) 
\end{equation}
depends only on the spatial coordinate $x$ and, via $X$, also on the mass parameter $m_0$.

To give one example we choose $V(X)=1/X^2$, obtaining the solution above with
\begin{equation}
\Phi(x) = -\frac{m_0}{x}
    \label{eq:NCex}
\end{equation}
which is just the Newton-potential in three spatial dimensions for an object of mass $m_0$. This shows that one can obtain higher-dimensional Newton potentials by choosing $V(X)$ suitably, which is again identical to how things work in usual 2d dilaton gravity.

Along the same lines we also propose general Carrollian dilaton gravity 
\begin{equation}
\lagr_{\tiny \textrm{Car-dil}}= X\,\dd \omega +X_\Ht\, (\dd \tau+ \omega \wedge e)+ X_\Pt\, \dd e + V(X,\,X_\Ht)\,\tau \wedge e\,.
\label{eq:cardeal}
\end{equation}
Here we eliminated the $\mathfrak{u}(1)$ field by a redefinition of $\Ht$. The $X$-dependence of the new potential term $V$ determines the curvature of the geometry with linear $X$-dependence corresponding to constant curvature. A non-trivial $X_\Ht$-dependence on the other hand leads to non-vanishing torsion. 
The discussion of equations of motion, constant and linear dilaton sectors is analogous to the NC case above.
\begin{align}
    &\textrm{constant\;dilaton}: & X_\Ht &=0 & X &= X^c & X_\Pt &= X_\Pt^c \\
    &\textrm{linear\;dilaton}: & X_\Ht &= X_\Ht(t,\,x) & X &= X(t,\,x) & X_\Pt &= X_\Pt(t,\,x)
\end{align}
The constant $X^c$ is again determined as roots of the potential,
    $V(X^c,\,0) = 0$, and curvature is again constant in the constant dilaton sector, $\ast\dd\omega=-\partial_XV$. 

\section{Metric BF theories and their limits}
\label{sec:limit}

For metric BF theories we demand the existence of an invariant metric
on the gauge algebra. In that case taking limits, especially of the
invariant metric of the Lie algebra, is more subtle since the
contracted Lie algebras do not necessarily inherit the non-degeneracy
of the metric. Since many interesting theories are based on metric Lie
algebras and this additional structure plays a r\^ole for our setup of
boundary conditions, we explain first when the limit has the chance to
lead to another metric BF theory. Then we show how the algebras and
limits can be generalized to also obtain Poincar\'e, Carrollian and
Galilean BF theories and observe the relation of the latter to NC
theory. Before discussing boundary conditions in the next section we
provide a summary of the various theories we have unveiled at the end
of this section.

\subsection{Metrics and limits}
\label{sec:why-they-do}

Starting point is the decomposed (A)dS algebra \eqref{eq:AdScov}
  \begin{align}
  \label{eq:adsstart}
  [\Bt,\Ht] &= \mp \tc^{2}  \Pt &  [\Bt,\Pt] &= \cc^{2} \Ht &  [\Ht,\Pt] &= -\lc  \Bt
  \end{align}
with the invariant metric
  \begin{align}
  \label{eq:adsmetrst}
 \langle \Bt,\Bt \rangle&= \pm \mu \cc^{2} \tc^{2}  &
 \langle \Ht,\Ht \rangle&=  \pm \mu \tc^{2} \lc&
 \langle \Pt,\Pt \rangle&=  \mu \cc^{2}\lc
\end{align}
parametrized such that any limit of the Lie algebra is well defined on
the level of the Lie algebra and metric. Taking either the flat
($\lc \to 0$), Galilean ($\cc =\frac{1}{c}\to 0$) or
Carrollian ($\tc \to 0$) limit of \eqref{eq:adsmetrst} leads
to a degenerate metric. In case we want to end up with a metric BF
theory there exists a necessary condition
\begin{equation}
  \label{eq:neccriterion}
  \textrm{dim}\,\mathfrak{g}=\textrm{dim} \,[\mathfrak{g},\mathfrak{g}]+\textrm{dim} \,Z(\mathfrak{g}) 
\end{equation}
for the existence of a metric on a Lie algebra $\mathfrak{g}$ where $Z(\mathfrak{g})$ is the center of the Lie algebra (see, e.g.,
Section 3.2 in~\cite{Matulich:2019cdo}). As long as we do not take any
limit, the center is trivial and $3 = 3+0$. However, taking any limit
reduces $\textrm{dim} \,[\mathfrak{g},\mathfrak{g}]$ by one without
adding any element to the center, i.e., $3\neq 2+0$ and no invariant
metric is possible. The addition of a nontrivial central element adds
one element to the dimension of the Lie algebra (on the left hand
side), but two on the right hand side, balancing the equation again,
$3+1=(2+1)+(0+1)$.

In the following sections we show that the addition of central extensions is also sufficient to
equip the Lie algebras and theories with an invariant metric and show
how these theories can be obtained from a limit.
Another option to obtain algebras with a non-degenerate invariant metric is based
on so called coadjoint Lie algebras as discussed in
Appendix~\ref{sec:coadjoint-theories}.

\subsection{Flat space dilaton gravity}
\label{sec:taking-flat-limit}

A metric BF formulation for the Poincar\'e algebra is only possible
when the algebra is suitably centrally extended. We show now how to
obtain the resulting theory as a limit of (A)dS at the level of the
action.

We start with the (A)dS algebra \eqref{eq:AdScov} and add an
additional generator $\Mt$ such that the algebra is given by
\begin{align}
  [\Bt,\Pt_{a}] &= -\epsilon\indices{_{a}^{b}} \Pt_{b} &
  [\Pt_{a},\Pt_{b}] &=  \epsilon_{ab}(-\lc \Bt+ 
  \Mt) \, .
\end{align}
At this point $\Mt$ is still a trivial central extension, but it is
introduced such that after the flat limit $\lc \to 0$ it is nontrivial
and leads to an invariant metric for (centrally extended) Poincar\'e.
The well-defined limit on the Lie algebra implies upon substitution of
the fields a limit of the equations of motion.

It remains to show that we can also take the limit at the level of the
action, for which we introduce the shifted invariant metric (c.f.\
\eqref{eq:adsmetrfixed})
\begin{subequations}
  \begin{align}
 \langle \Bt,\Bt \rangle&=  \frac{\mu+\mu_{2}
 }{\lc}   &
 \langle \Pt_{a},\Pt_{b} \rangle&=  \mu  \eta_{ab}  \\
 \langle \Mt,\Mt \rangle&= \mu_{2}\lc &
 \langle \Bt,\Mt \rangle&=  \mu_{2} 
\end{align}
\end{subequations}
where we used $\mu \mapsto \frac{\mu}{\lc}$. We now take the flat
limit $\lc \to 0$, assuming $\chi=(\mu+\mu_2)/\lc$ remains finite,
yielding
\begin{align}
  \label{eq:flatcov}
  [\Bt,\Pt_{a}] &= -\epsilon\indices{_{a}^{b}} \Pt_{b} &
  [\Pt_{a},\Pt_{b}] &=  \epsilon_{ab} \Mt
\end{align}
with invariant metric
\begin{align}
 \langle \Pt_{a},\Pt_{b} \rangle&=  \mu  \eta_{ab} & \langle \Bt,\Mt \rangle&= - \mu & \langle \Bt,\Bt \rangle&= \chi \, .
\end{align}  
This is the (non-semisimple) extended Poincar\'e algebra with
non-degenerate metric for $\mu \neq 0$. This means we can write down a
metric BF theory~\eqref{eq:mBF}, which yields the CGHS
model~\cite{Callan:1992rs} in the formulation of Cangemi and
Jackiw~\cite{Cangemi:1992bj} (see also
\cite{Verlinde:1991rf,Jackiw:1992bw}).

The algebra \eqref{eq:flatcov} has a higher dimensional
generalization, called Maxwell algebra, which emerges in the study of
particles in classical homogeneous electromagnetic
fields~\cite{Schrader:1972zd,Bacry:1970ye}. This is also true for the
extended Poincar\'e (Euclidean) algebra which arise when considering a
charged particle in 1+1 dimensions in a constant electric (magnetic)
field. From the point of view of metric Lie algebras the Maxwell
algebra is the natural metric generalization of
Poincar\'e~\cite{Matulich:2019cdo}.

The Carrollian and Galilean limits can be done analogously to the
flat limit above. They lead to (A)dS Carroll and (A)dS Galilei
theories, respectively, and are related to the CGHS model via
geometric dualities as summarized in Table~\ref{tab:Liealg2}.

\subsection{Summary}
\label{sec:summary}

We have summarized interesting homogeneous spaces and Lie algebras in
Table~\ref{tab:Liealg2}, which is best read together with
Figure~\ref{fig:cube}.

The table starts by providing the necessary information to construct
the theories based on the simple Lie algebras
$\mathfrak{sl}(2,\R) \simeq \mathfrak{so}(2,1) \simeq
\mathfrak{so}(1,2)$ and $\mathfrak{so}(3)$. They encompass the well
known (A)dS BF theories, their Euclidean cousins and the
light cone BF theory of Section \ref{sec:bf-light-cone}.
Table~\ref{tab:Liealg2} makes explicit that all, but the sphere, are
based on the same Lie algebra; however, they differ as homogeneous
spaces (under the additional assumption that we disallow an exchange
of time and space) and therefore in their geometric and physical
interpretation, as described in Section \ref{sec:choice-spacetime}.

Taking either one of the flat/Carrollian/Galilean limits of the (A)dS
theory we arrive at Poincar\'e/(A)dS Carrollian/(A)dS Galilean
theories. These BF theories are based on Lie algebras that do not
admit an invariant metric, see Section \ref{sec:limits-bf-theories}.
They allow for one nontrivial central extension that renders the
invariant metric non-degenerate. These (centrally) extended theories
can also be obtained from a limit, shown in Section
\ref{sec:taking-flat-limit}.

Taking a second limit, e.g., first the flat and then the Carrollian,
leads to the Carroll, Galilei and para-Galilei theories. For these
theories doubly centrally extending leads from a degenerate to a
non-degenerate invariant metric. Using the procedure described in
Section~\ref{sec:taking-flat-limit} one can show that all the
theories below the first horizontal dividing line can be
obtained by contraction and taking quotients starting from the parent
theory
$\mathfrak{sl}(2,\R) \oplus \mathfrak{u}(1)\oplus \mathfrak{u}(1)$.
The two $\mathfrak{u}(1)$'s correspond to central extensions that are trivial
before taking limits but become non-trivial afterwards.
Nevertheless, the theory still describes a standard NC/Carroll structure since the additional field associated to the $\mathfrak{u}(1)$ generator $\Zt$, introduced in order to have a non-degenerate metric, decouples on-shell.

Taking all three limits leads to the static case, yielding an abelian
algebra, which always allows for an invariant metric, and fulfills the
necessary condition \eqref{eq:neccriterion} since $3=0+3$. For the
static and the para-Galilei case the group action of the boosts on the
homogeneous space is trivial, i.e., they do not act at all and leave
points unaltered. It is then natural to quotient by them, leading to
an Aristotelian algebra that is abelian in 1+1 dimensions.

Most of our theories arise from limits of centrally extended (A)dS,
but there are further interesting spaces on which one can base
Lifshitz, Schr\"odinger, 1/c expanded Poincar\'e, and coadjoint
theories, that we discuss in Appendix~\ref{sec:lifshitz-schrodinger}
and \ref{sec:coadjoint-theories}. Yet another generalization are
theories based on the remaining kinematical homogeneous spaces that do
not follow from a limit (we do not discuss the cases of torsional
Galilean and S17-S20 of Table~1 in~\cite{Figueroa-OFarrill:2018ilb}).
    
\newcommand{\Hei}{\textrm{Heisenberg}}
\newcommand{\iso}{\mathfrak{iso}}
\begin{table}[ht]
  \centering
   \resizebox{\linewidth}{!}{
  \begin{tabular}{L L |L L|  L L L | C | L L L L L L}\toprule
                                                                            & \multicolumn{1}{c|}{Limit}       & \multicolumn{5}{c|}{}                                                & \multicolumn{5}{c}{}                                                                                                                                                                                                                                                                        \\
    \multirow{-2}{*}{\text{Homogeneous space}}                              & \multicolumn{1}{c|}{($\to 0$)}   & \multicolumn{5}{c|}{\multirow{-2}{*}{Nonzero commutation relations}} & \text{ND?}                   & \multicolumn{4}{c}{\multirow{-2}{*}{Invariant metric}}                                                                                                                                                                                                                                                                        \\\midrule
    \text{Sphere}                                                           &                                  & \so                                                                  & \textrm{IX}                  & [\Bt,\Ht] =-\Pt                     & [\Bt,\Pt] =\Ht                      & [\Ht,\Pt] =- \Bt                     & \yes                   & \langle \Bt,\Bt \rangle= \mu            & \langle \Ht,\Ht \rangle= \mu                           & \langle \Pt,\Pt \rangle=  \mu \\
    \rowcolor{red!10} \text{de Sitter (dS)}                                 &                                  & \So(2,1)                                                             & \textrm{VIII}                & [\Bt,\Ht] =\Pt                      & [\Bt,\Pt] =\Ht                      & [\Ht,\Pt] = -\Bt                     & \yes                   & \langle \Bt,\Bt \rangle= -\mu           & \langle \Ht,\Ht \rangle= - \mu                         & \langle \Pt,\Pt \rangle=  \mu                                    \\
    \text{Hyperbolic}                                                       &                                  & \So(2,1)                                                             & \textrm{VIII}                & [\Bt,\Ht] =-\Pt                     & [\Bt,\Pt] =\Ht                      & [\Ht,\Pt] = \Bt                      & \yes                   & \langle \Bt,\Bt \rangle= \mu            & \langle \Ht,\Ht \rangle= - \mu                         & \langle \Pt,\Pt \rangle= - \mu                                   \\
    \rowcolor{red!10} \text{Anti-de Sitter (AdS)}                           &                                  & \So(1,2)                                                             & \textrm{VIII}                & [\Bt,\Ht] =\Pt                      & [\Bt,\Pt] =\Ht                      & [\Ht,\Pt] = \Bt                      & \yes                   & \langle \Bt,\Bt \rangle= -\mu           & \langle \Ht,\Ht \rangle=  \mu                          & \langle \Pt,\Pt \rangle= - \mu \\
    \text{Light cone}                                       &                                  & \slR                                                                 & \textrm{VIII}                & [\Bt,\Ht] =-\Bt                     & [\Bt,\Pt] =\Ht                      & [\Ht,\Pt] = - \Pt                    & \yes                   & \langle \Bt,\Pt \rangle= \mu            & \langle \Ht,\Ht \rangle= \mu                                                            \\ \midrule
    \text{Euclidean}                                                        & \lc                              & \iso(2)                                                              & \eu                          & [\Bt,\Ht] =-\Pt                     & [\Bt,\Pt] =\Ht                      &                                      & \no                    & \langle \Bt,\Bt \rangle= \mu            &                                                        & \\
    \rowcolor{green!10}      \text{Extended Euclidean}                      & \lc                              & \iso(2)^c                                                            & \euc                         & [\Bt,\Ht] =-\Pt                     & [\Bt,\Pt] =\Ht                      & [\Ht,\Pt] = \Mt                      & \yes                   & \langle \Bt , \Bt \rangle= \chi         & \langle \Ht,\Ht \rangle= \mu                           & \langle \Pt,\Pt \rangle=  \mu & \langle \Bt ,\Mt \rangle=  - \mu \\
    \text{dS-Carroll}                                                       & \tc                              & \iso(2)                                                              & \eu                          &                                     & [\Bt,\Pt] =\Ht                      & [\Ht,\Pt] =- \Bt                     & \no                    &                                         &                                                        & \langle \Pt,\Pt \rangle=  \mu                                    \\ 
    \rowcolor{green!10}      \text{Extended dS-Carroll}                     & \tc                              & \iso(2)^c                                                            & \euc                         & [\Bt,\Ht] =\Mt                      & [\Bt,\Pt] =\Ht                      & [\Ht,\Pt] =-\Bt                      & \yes                   & \langle \Bt , \Bt \rangle= \mu          & \langle \Ht,\Ht \rangle= \mu                           & \langle \Pt,\Pt \rangle= \chi & \langle \Pt ,\Mt \rangle= - \mu \\ 
    \text{AdS--Galilei}                                                      & \cc                              & \iso(2)                                                              & \eu                          & [\Bt,\Ht] =\Pt                      &                                     & [\Ht,\Pt] = \Bt                      & \no                    &                                         & \langle \Ht,\Ht \rangle=  \mu                          & \\ 
    \rowcolor{green!10}      \text{Extended AdS--Galilei}                    & \cc                              & \iso(2)^c                                                            & \euc                         & [\Bt,\Ht] =\Pt                      & [\Bt,\Pt] =\Mt                      & [\Ht,\Pt] = \Bt                      & \yes                   & \langle \Bt , \Bt \rangle  = \mu        & \langle \Ht,\Ht \rangle= \chi                          & \langle \Pt,\Pt \rangle= \mu  & \langle \Ht ,\Mt \rangle= - \mu \\  \midrule
    \text{Poincar\'e}                                                         & \lc                              & \iso(1,1)                                                            & \poi                         & [\Bt,\Ht] =\Pt                      & [\Bt,\Pt] =\Ht                      &                                      & \no                    & \langle \Bt,\Bt \rangle= -\mu           &                                                        & \\ 
    \rowcolor{green!10}      \text{Extended Poincar\'e}                       & \lc                              & \iso(1,1)^c                                                          & \poic                        & [\Bt,\Ht] =\Pt                      & [\Bt,\Pt] =\Ht                      & [\Ht,\Pt] = \Mt                      & \yes                   & \langle \Bt , \Bt \rangle= \chi         & \langle \Ht,\Ht \rangle= - \mu                         & \langle \Pt,\Pt \rangle=  \mu & \langle \Bt ,\Mt \rangle=   \mu  \\
    \text{AdS--Carroll}                                                      & \tc                              & \iso(1,1)                                                            & \poi                         &                                     & [\Bt,\Pt] =\Ht                      & [\Ht,\Pt] = \Bt                      & \no                    &                                         &                                                        & \langle \Pt,\Pt \rangle= - \mu                                   \\ 
    \rowcolor{green!10}      \text{Extended AdS--Carroll}                    & \tc                              & \iso(1,1)^c                                                          & \poic                        & [\Bt,\Ht] =\Mt                      & [\Bt,\Pt] =\Ht                      & [\Ht,\Pt] = \Bt                      & \yes                   & \langle \Bt , \Bt \rangle= \mu          & \langle \Ht,\Ht \rangle= - \mu                         & \langle \Pt,\Pt \rangle= \chi & \langle \Pt ,\Mt \rangle=  \mu   \\ 
    \text{dS-Galilei}                                                       & \cc                              & \iso(1,1)                                                            & \poi                         & [\Bt,\Ht] =\Pt                      &                                     & [\Ht,\Pt] = - \Bt                    & \no                    &                                         & \langle \Ht,\Ht \rangle= - \mu                         & \\ 
    \rowcolor{green!10}      \text{Extended dS-Galilei}                     & \cc                              & \iso(1,1)^c                                                          & \poic                        & [\Bt,\Ht] =\Pt                      & [\Bt,\Pt] =\Mt                      & [\Ht,\Pt] = - \Bt                    & \yes                   & \langle \Bt , \Bt \rangle  = \mu        & \langle \Ht,\Ht \rangle= \chi                          & \langle \Pt,\Pt \rangle= -\mu & \langle \Ht ,\Mt \rangle= \mu \\ \midrule 
    \text{Carroll}                                                          & \tc, \lc                         & \Hei                                                                 & \hei                         &                                     & [\Bt,\Pt] =\Ht                      &                                      & \no                    & \langle \Bt,\Bt \rangle= \chi_{\Bt}     & \langle \Bt,\Pt \rangle= \chi_{\Bt \Pt}                & \langle \Pt,\Pt \rangle= \chi_{\Pt}                    \\ 
    \rowcolor{blue!10}                                                      &                                  &                                                                      &                              &                                     &                                     &                                      &                        & \langle \Ht,\Ht \rangle=  \mu           & \langle \Bt,\Zt \rangle= - \mu                         & \langle \Pt,\Mt \rangle  = - \mu                       \\ 
    \rowcolor{blue!10}   \multirow{-2}{*}{Doubly extended Carroll}          & \multirow{-2}{*}{$\tc, \lc$}     & \multirow{-2}{*}{$\Hei^{cc}$}                                        & \multirow{-2}{*}{$\heicc$}   & \multirow{-2}{*}{$[\Bt,\Ht] =\Mt$}  & \multirow{-2}{*}{$[\Bt,\Pt] =\Ht$}  & \multirow{-2}{*}{$[\Ht,\Pt] = \Zt$}  & \multirow{-2}{*}{\yes} & \langle \Bt,\Bt \rangle  = \chi_{\Bt}   & \langle \Bt,\Pt \rangle               = \chi_{\Bt \Pt} & \langle \Pt,\Pt \rangle                   = \chi_{\Pt} \\ 
    \text{Galilei}                                                          & \cc, \lc                         & \Hei                                                                 & \hei                         & [\Bt,\Ht] =\Pt                      &                                     &                                      & \no                    & \langle \Bt,\Bt \rangle= \chi_{\Bt}     & \langle \Bt,\Ht \rangle= \chi_{\Bt \Ht}                & \langle \Ht,\Ht \rangle= \chi_{\Ht}     \\ 
    \rowcolor{blue!10}                                                      &                                  &                                                                      &                              &                                     &                                     &                                      &                        & \langle \Pt,\Pt \rangle=\mu             & \langle \Bt,\Zt \rangle= \mu                           & \langle \Ht,\Mt \rangle  = - \mu        \\ 
    \rowcolor{blue!10}       \multirow{-2}{*}{Doubly extended Galilei}      & \multirow{-2}{*}{$\cc, \lc$}     & \multirow{-2}{*}{$\Hei^{cc}$}                                        & \multirow{-2}{*}{$\heicc$}   & \multirow{-2}{*}{$[\Bt,\Ht] = \Pt$} & \multirow{-2}{*}{$[\Bt,\Pt] = \Mt$} & \multirow{-2}{*}{$[\Ht,\Pt] =  \Zt$} & \multirow{-2}{*}{\yes} & \langle \Bt,\Bt \rangle= \chi_{\Bt}     & \langle \Bt,\Ht \rangle= \chi_{\Bt \Ht}                & \langle \Ht,\Ht \rangle= \chi_{\Ht}     \\ 
    \text{Para-Galilei}                                                     & \cc, \tc                         & \Hei                                                                 & \hei                         &                                     &                                     & [\Ht,\Pt]=\Bt                        & \no                    & \langle \Ht,\Ht \rangle= \chi_{\Ht}     & \langle \Ht,\Pt \rangle= \chi_{\Ht \Pt}                & \langle \Pt,\Pt \rangle= \chi_{\Pt}     \\ 
    \rowcolor{blue!10}                                                      &                                  &                                                                      &                              &                                     &                                     &                                      &                        & \langle \Bt,\Bt \rangle= \mu            & \langle \Ht,\Zt \rangle= -\mu                          & \langle \Pt,\Mt \rangle= \mu            \\ 
    \rowcolor{blue!10}       \multirow{-2}{*}{Doubly extended para-Galilei} & \multirow{-2}{*}{$\cc, \tc$}     & \multirow{-2}{*}{$\Hei^{cc}$}                                        & \multirow{-2}{*}{$\heicc$}   & \multirow{-2}{*}{$[\Bt,\Ht] = \Mt$} & \multirow{-2}{*}{$[\Bt,\Pt] = \Zt$} & \multirow{-2}{*}{$[\Ht,\Pt] =  \Bt$} & \multirow{-2}{*}{\yes} & \langle \Ht,\Ht \rangle= \chi_{\Ht}     & \langle \Ht,\Pt \rangle= \chi_{\Ht \Pt}                & \langle \Pt,\Pt \rangle= \chi_{\Pt}     \\ \midrule 
                                                                            &                                  &                                                                      &                              &                                     &                                     &                                      &                        & \langle \Bt,\Bt \rangle= \mu_{\Bt}      & \langle \Ht,\Ht \rangle= \mu_{\Ht}                     & \langle \Pt,\Pt \rangle= \mu_{\Pt}      \\ 
         \multirow{-2}{*}{Static}                                           & \multirow{-2}{*}{$\cc, \tc,\lc$} & \multirow{-2}{*}{Abelian}                                            & \multirow{-2}{*}{\textrm{I}} &                                     &                                     &                                      & \multirow{-2}{*}{\yes} & \langle \Bt,\Ht \rangle= \chi_{\Bt \Ht} & \langle \Bt,\Pt \rangle= \chi_{\Bt \Pt}                & \langle \Ht,\Pt \rangle= \chi_{\Ht \Pt} \\  
    \bottomrule
    \end{tabular}
  }
  \caption[forparagraphs]{Overview of homogeneous spaces/Klein pairs
    $(\mathfrak{g},\Bt)$, underlying Lie algebras $\mathfrak{g}$ and
    their invariant metric (`ND?' indicates Non-Degeneracy). The real Lie algebras are indicated by
    their names and according to Bianchi's
    classification~\cite{Bianchi,MR1900159}. Superscripts $c$ denote
    nontrivial central extensions. Rows in red are Lorentzian versions
    of the previous row, rows in green (blue) are centrally extended
    once (twice) and arise as limits of
    $\mathfrak{sl}(2,\R) \oplus \mathfrak{u}(1)$
    ($\mathfrak{sl}(2,\R) \oplus \mathfrak{u}(1)\oplus
    \mathfrak{u}(1)$), see Section \ref{sec:taking-flat-limit}.
  }
    \label{tab:Liealg2}
\end{table}

\section{Boundary actions of kinematical BF theories}
\label{sec:boundary}

In this section we discuss boundary actions associated with the
kinematical BF theories introduced in the previous sections,
restricted to metric BF theories.

\subsection{Particle on group manifold}
\label{sec:particle}

In the case of Chern--Simons theories in three dimensions it is
well-known that these theories reduce to Wess--Zumino--Witten (WZW) models on
manifolds with boundaries \cite{Witten:1988hf,Elitzur:1989nr}. In a
similar way, it can be shown that BF theories with a particular choice
of boundary condition reduce to the action of a particle on the group
manifold of the chosen gauge group. 

The variation of the action of a metric BF theory \eqref{eq:mBF} on a
manifold with boundary reads
\begin{equation}
  \label{eq:1}
  \delta I
  =\int_{\mathcal{M}}\left(\langle \delta \mathcal{X},F\rangle-\langle \dd \mathcal{X}+[A,\mathcal{X}],\delta A\rangle\right)
  +\int_{\partial{\mathcal{M}}}\langle \mathcal{X},\delta A\rangle\,.
\end{equation}
A possible choice of boundary conditions is to take Dirichlet boundary
conditions on $A$. With this choice no further boundary term is needed
for a well-defined variational principle and the theory is topological
without any dynamics on the boundary. If we are to interpret the
connection components as zweibein and spin-connection for a
gravitational theory, putting Dirichlet boundary conditions on all
components of $A$ is in general too strict.\footnote{This is related
  to the imposition of either Dirichlet or Neumann conditions on the
  metric (and not both).} Instead of Dirichlet boundary conditions we impose
\begin{equation}
  \label{eq:groupbc}
  \mathcal{X}\,df |_{\partial \mathcal{M}}=A|_{\partial \mathcal{M}}\, .
\end{equation}
Note that we implicitly used the invariant metric here in order to write
both $\mathcal{X}$ and $A$ as elements of the Lie algebra.
The one-form $df$ is assumed to be fixed on the boundary
\cite{Mertens:2018fds,Gonzalez:2018enk,Saad:2019lba,Kapec:2019ecr}.\footnote{In
  \cite{Kapec:2019ecr} this boundary condition was interpreted as a
  Yang--Mills theory with a position-dependent coupling constant that
  is localized near the boundary.} Imposing this boundary condition
requires the addition of a boundary term to \eqref{eq:mBF} such that
the full action reads
\begin{equation}
  \label{eq:fullaction}
  I[A,\mathcal{X}]=\int_{\mathcal{M}}\langle \mathcal{X},F\rangle-\frac{1}{2}\int_{\partial\mathcal{M}}\dd f \langle \mathcal{X},\mathcal{X}\rangle\,.
\end{equation}
The integrand of the boundary term is recognized as the quadratic Casimir 
\begin{equation}
    \label{eq:quadCas}
    C=\langle \mathcal{X},\mathcal{X}\rangle\,.
\end{equation}
that is conserved on-shell using the right hand side of~\eqref{eq:EOM}.

We introduce a coordinate system
$(\rho,\,\timet)$ with $\timet$ being the coordinate along the boundary
that is located at $\rho\rightarrow\infty$. We assume that the
homogeneous space on which the gravitational theory is defined has
only one boundary component, the topology of which can be either a
circle or a line.

Starting from \eqref{eq:fullaction} one can integrate out the dilaton
field enforcing the constraint $F=0$ which is solved locally by
\begin{equation}
  \label{eq:2}
  A=\tilde{g}^{-1}\dd \tilde{g}
\end{equation}
where $\tilde{g}$ are elements of the gauge group of the BF theory. In
order to simplify the discussion we assume that close to the boundary
the group element $\tilde{g}$ factorizes
\begin{equation}
  \label{eq:factorg}
  \tilde{g}= g(\timet) b(\rho)\,.
\end{equation}

Using the boundary condition \eqref{eq:groupbc} and assuming the
orientation $(\rho,\timet)$, the action \eqref{eq:fullaction} becomes
\begin{equation}
  \label{eq:groupaction}
  I[g]=-\frac{1}{2}\int_{\partial \mathcal{M}}\dd \timet (\partial_{\timet}f)^{-1}\langle g^{-1}\partial_\timet g,g^{-1}\partial_\timet g\rangle
\end{equation}
which is the action for a particle moving on the group manifold of
$G$. The gauge modes $g$ thus become physical at the boundary due to
the explicit breaking of gauge invariance by the boundary condition
\eqref{eq:groupbc}. This action is invariant under two copies of the
global symmetry group, i.e., under the transformation
\begin{equation}
  \label{eq:26}
  g\mapsto \tilde{h} g h \qquad \text{with constant }  \tilde{h},h \in G \,
\end{equation}
with corresponding left and right charges $\tilde{Q}$ and $Q$
\begin{equation}
  \label{eq:32}
  \tilde{Q}_{\tilde{A}}=(\partial_\timet f)^{-1}\langle \tilde{A}, \partial_\timet g g^{-1}\rangle\,\qquad Q_A=(\partial_\timet f)^{-1}\langle A, g^{-1}\partial_\timet g\rangle\,\qquad  \tilde{A}, A \in \mathfrak{g}
\end{equation}
where  $\tilde{A}$ and $A$ are the Lie algebra generators of the corresponding group elements $\tilde{h}$ and $h$, respectively.
The right charges have the Poisson bracket
\begin{equation}
  \label{eq:37}
\{Q_A,Q_B\}=Q_{[A,B]}
\end{equation}
with a similar expression for the left charges. Note, however, that
the left transformations $\tilde{h}$ correspond to a redundancy in our
reduction \eqref{eq:2} and should therefore be thought of as being
gauged, i.e., all the left charges should be set to zero on-shell.
This requirement is extrinsic to the action \eqref{eq:groupaction} and
is a consequence of arriving at this action from a 2d
bulk action.\footnote{Compare this to the three dimensional cases
  \cite{Cotler:2018zff,Merbis:2019wgk} where the two-dimensional
  boundary actions also inherit a gauge symmetry corresponding to the
  global symmetry group.} Taken together, the above action is the
one-dimensional equivalent to the WZW action appearing at the boundary
of a CS theory in three dimensions.

In deriving the action we have only used the boundary condition
\eqref{eq:groupbc} which can be imposed for any metric BF theory.
But this is not the end of the story as ultimately we want to
interpret these BF theories as (non- or ultra-relativistic) theories of
gravity where one might impose additional asymptotic boundary conditions on the
metric or, equivalently, on vielbein and spin-connection. In other words, we are looking for a (Hamiltonian) reduction of the boundary action \eqref{eq:groupaction}.

\subsection{Hamiltonian reduction of boundary action}
\label{sec:constr-part-acti}

For many applications we are not interested in the loosest set of boundary conditions, but rather impose (physically or geometrically motivated) restrictions on the fields. A famous example is the Drinfeld--Sokolov reduction of the $\mathfrak{sl}(2,\R)$ current algebra to the Virasoro algebra, which applied to AdS$_3$ gravity yields Brown--Henneaux boundary conditions \cite{Brown:1986nw}. For AdS$_3$ gravity several inequivalent boundary conditions were identified, e.g., \cite{Brown:1986nw,Compere:2013bya,Troessaert:2013fma,Grumiller:2016pqb}. From the point of view of the boundary WZW model these boundary conditions act as current constraints \cite{Coussaert:1995zp}. Following this procedure one finds the Alekseev--Shatashvili action as boundary theory for AdS$_3$ with Brown--Henneaux boundary conditions \cite{Cotler:2018zff} or the $\textrm{BMS}_3$ geometric action \cite{Merbis:2019wgk} for the boundary conditions of \cite{Barnich:2006av} on three-dimensional flat space. Both of these theories have a flavor of hydrodynamics in the sense that their dynamics is nothing but the conservation of the stress-energy tensor (or its Carrollian analogue in the flat space case). As reviewed below, the Schwarzian action for $\textrm{AdS}_2$ can be understood in a similar way.
 
Boundary conditions on spin-connection and vielbein translate to conditions on the gauge-connection $A$ and by the flatness condition \eqref{eq:2} to constraints on $g^{-1}\partial_\timet g$. Assuming that the boundary conditions are consistent with the algebraic structure of the particle action, we can view the former as constraints on the right charges $Q_A$.\footnote{In the case of WZW models based on simple algebras, the question of consistent sets of constraints has been analyzed in detail in \cite{Feher:1992ed}.} The structure of the algebra puts restrictions on the set of possible consistent constraints and thus possible boundary conditions. In what follows we make some specific assumptions about the boundary conditions that could be (and for some applications have to be) relaxed; however, they will be useful for the two examples that we provide in Section \ref{sec:schwarzianBF}.

Let $Q_A$ be the set of right charges of the action
\eqref{eq:groupaction} based on the Lie algebra $\mathfrak{g}$. A
generic constraint compatible with the algebra structure on
$g^{-1}\partial_\timet g$ has the form
\begin{equation}
  \label{eq:39}
  \Phi_\gamma\equiv \langle \gamma, g^{-1}\partial_\timet g\rangle- \langle \gamma, K\rangle=0\qquad \qquad\gamma, K\in \mathfrak{g}\,.
\end{equation}
The constraint thus sets the charge $Q_\gamma$ on-shell to some value determined by the fixed algebra element $K$. (By fixed we imply that $K$ has vanishing Poisson brackets with all functions on phase space; we assume the same for $\gamma$).

Using the Poisson bracket for the right charges \eqref{eq:37} we calculate the Poisson bracket of the constraints 
\begin{equation}
  \label{eq:40}
  (\partial_\timet f)^{-1}\{\Phi_{\gamma_1},\Phi_{\gamma_2}\}=\Phi_{[\gamma_1,\gamma_2]}+\langle [\gamma_1,\gamma_2],K\rangle\,.
\end{equation}

We note first that a new constraint is generated on the right hand side of \eqref{eq:40} unless all the generators belong to a subalgebra, i.e., $\gamma_1,\gamma_2\in \Gamma$ where $[\Gamma,\Gamma]\subset \Gamma$. The nature of the constraints is now determined by the second term. If $K\in \Gamma^\perp$ or the subalgebra is abelian, the constraints are first-class.\footnote{%
We use here the standard terminology to refer to constraints whose Poisson brackets vanish on-shell as \emph{first class}. Constraints with Poisson brackets that do not vanish on the constraint surface are called \emph{second class}. From this follows immediately that second class constraints come always in even numbers.} 
First-class constraints are the hallmark of gauge symmetry in a system and require additional constraints in the form of gauge-fixing conditions that render the first-class constraints second-class. For the type of boundary conditions we are currently interested in, we do not want to introduce (further) gauge symmetries. Therefore, we demand that the system in \eqref{eq:40} be second-class. 

This means we have found a way to determine consistent boundary conditions compatible with the algebraic structure of the boundary action \eqref{eq:groupaction} by looking for \emph{even-dimensional non-abelian subalgebras $\Gamma$ of the gauge algebra $\mathfrak{g}$. The constant element $K$ is subsequently chosen such that $\langle K,\Gamma\rangle\neq 0$.}

In summary, boundary conditions on the fields of the BF theory compatible with the universal boundary condition \eqref{eq:groupbc} lead to the action \eqref{eq:groupaction} together with a system of second class constraints. The latter can be solved directly in the action \cite{Henneaux:1992ig}.

In the above we spelled out a purely algebraic way to arrive at consistent boundary condition for a metric BF theory. This is usually not the way boundary conditions for gravitational theories are conceived. Rather, one starts from a bulk perspective and chooses boundary conditions such that they allow an interesting class of bulk geometries as solutions. This is indeed the case for the BF theory on $\textrm{AdS}_2$, i.e., the JT model in the second order perspective, and we will see below that these boundary conditions satisfy the above criteria. But the purely algebraic point of view presented here allows also to find boundary conditions for BF theories with Carrollian/Galilean interpretation where the geometric picture is often not as clear as in the relativistic case. In the next section we apply this procedure to two examples.

\section{Schwarzian-like theories}
\label{sec:schwarzianBF}

In this section we consider metric BF theories based on the various (1+1)-dimensional kinematical algebras and look for boundary conditions, as discussed in the previous section, that reduce the particle action \eqref{eq:groupaction} to Schwarzian-like actions.

To show how our proposal in Section \ref{sec:boundary} works, we first review the construction of the Schwarzian action for a BF theory based on the symmetry algebra of Euclidean $\textrm{AdS}_2$ in Section \ref{sec:schwarzianAdS}. The main new example is the construction of the boundary action for (extended) AdS--Carroll$_2$ in Section \ref{sec:AdSCaction}. Other kinematical algebras where our procedure does not work without modifications are briefly mentioned in Section \ref{sec:remainingalg}. 

\subsection{Euclidean \texorpdfstring{$\textrm{AdS}_2$}{AdS2}/Hyperbolic plane}
\label{sec:schwarzianAdS}

The discussion in Section \ref{sec:constr-part-acti} instructs to look for two-dimensional non-abelian subalgebras of the Euclidean AdS algebra. In the $\Ht,\Bt,\Pt$ basis there is no obvious choice, but changing to the $\Lt_+,\Lt_0,\Lt_-$ basis
\begin{equation}
  \label{eq:6}
  \Lt_+=\Bt + \Ht\qquad\qquad \Lt_-=\Bt-\Ht\qquad\qquad \Lt_0=\Pt
\end{equation}
with commutation relations explicitly given by \eqref{eq:sl2r}, we find the two choices $\Gamma=\{\Lt_+,\Lt_0,\}$ and $\Gamma=\{\Lt_-,\Lt_0\}$. Taking the latter pair (the other one leads to the same conclusions upon redefinition of the radial coordinate) we find from \eqref{eq:40} that the fixed element $K$ has to be proportional to $\Lt_+$ in order to have non-zero inner product with $\Lt_-$. It can be shifted by elements of $\Lt_0$ and $\Lt_-$ although the latter element has no influence on the constraints. Choosing $K=\alpha_+ \Lt_++\alpha_0 \Lt_0$ with $\alpha_+\neq 0$ the constraints \eqref{eq:39} read
\begin{equation}
  \label{eq:42}
\phi_-=\langle \Lt_-, g^{-1}\partial_\timet g\rangle-2\alpha_+\qquad\qquad \phi_0=\langle \Lt_0, g^{-1}\partial_\timet g\rangle+\alpha_0
\end{equation}
which yields
\begin{equation}
  \label{eq:8}
  g^{-1}\partial_\timet g|_0=\alpha_0 \qquad
  g^{-1}\partial_\timet g|_+=\alpha_+\,.
\end{equation}
Assuming that the radial dependence of the connection \eqref{eq:3} is completely captured by the group element $b=\exp(\rho\, \Lt_0)$, these constraints reduce to those implied by the well-known boundary conditions for the connection on Euclidean $\textrm{AdS}_2$ (hyperbolic in Table~\ref{tab:Liealg2}) for some particular choice of $\alpha_0$ and $\alpha_+$, that is ultimately inconsequential for the reduced action. In the first order formulation these boundary conditions in the highest-weight gauge (cf., e.g., \cite{Grumiller:2017qao,Gonzalez:2018enk}) are given by
\begin{align}
  \label{eq:3}
  A_\timet=e^{\rho}\Lt_++\mathcal{L}(\timet)e^{-\rho}\Lt_-
  \qquad
  A_\rho=\Pt
\end{align}
with the generators \eqref{eq:adsstart} and the cosmological constant set to unity. In the second order formulation these boundary conditions translate to the metric
\begin{equation}
  \label{eq:4}
  \dd s^2=\dd \rho^2+\frac{1}{4} \left(e^{\rho}- \mathcal{L}(\timet)e^{-\rho}\right)^2\dd \timet^2
\end{equation}
where $\mathcal{L}(\timet)$ is an arbitrary function and the dilaton field has the asymptotic behavior
\begin{equation}
  \label{eq:5}
  X=e^{\rho}\bar{X}+O(e^{-\rho})\,.
\end{equation}
We have therefore reconstructed the above boundary conditions on $\textrm{AdS}_2$ out of the algebraic considerations of Section \ref{sec:constr-part-acti}.

Under gauge transformations generated by $\tfrac12\,\lambda\Lt_++\lambda^0\Lt_0+\lambda^-\Lt_-$ preserving the form of $A_\timet$, the field $\mathcal{L}(\timet)$ transforms with an infinitesimal Schwarzian derivative,
\begin{equation}
  \label{eq:43}
  \delta_{\lambda}\mathcal{L}(\timet)=\lambda \mathcal{L}' + 2\lambda' \mathcal{L} + \lambda'''
\end{equation}
which corresponds to the transformation of a stress tensor under infinitesimal conformal transformations. The zero-mode of the function $\mathcal{L}$ is related to the mass of spacetime \cite{Grumiller:2017qao}.

We turn now to the derivation of the boundary action that follows from the particle action \eqref{eq:groupaction} upon introducing the constraints implied by the above boundary conditions. Using Gauss parametrization\footnote{%
This parametrization is valid for $SO(2,1)=\mathrm{PSL}(2,\R)=\mathrm{SL}(2,\R)/\{\pm 1\}$ which is the group for which globally AdS$_{2}$ is single valued in the first order formulation.}
for the $\timet$-dependent group element
\begin{equation}
  g=e^{y \Lt_+}e^{y^0 \Lt_0}e^{y^- \Lt_-}
\end{equation}
the two constraints \eqref{eq:8} lead to two algebraic equations that can be solved in terms of the field $y$. Plugging the solutions into the action \eqref{eq:groupaction} yields
\begin{equation}
  \label{eq:9}
  I[y]=\frac{1}{2}\int \dd \timet (\partial_\timet f)^{-1} \{y,\timet\}\,.
\end{equation}
where $\{\cdot,\cdot\}$ denotes the Schwarzian derivative. Using the boundary condition \eqref{eq:groupbc} we can further relate $\partial_\timet f$ to the leading order of the dilaton so that we find
\begin{equation}
  \label{eq:10}
  I[y]=\bar{X}\int \dd \timet\, \{y,\timet\}\,
\end{equation}
thus reproducing the Schwarzian action. The action is invariant under the finite $\textrm{PSL}(2,\R)$ transformations
\begin{equation}
  \label{eq:27}
y\mapsto \frac{a y+b}{c y+d}\qquad \qquad ad-bc=1
\end{equation}
that correspond to group multiplication from the left, $\tilde{h}$ in \eqref{eq:26}. As argued in Section \ref{sec:schwarzianBF}, this transformation should be thought of as a gauge symmetry.

In terms of the group elements $g$ the function $\mathcal{L}(\timet)$ reads
\begin{equation}
  \label{eq:44}
 \mathcal{L}(\timet)=-\frac{\{y,\timet\}}{2}\,\,.
\end{equation}
Since the equation of motion of the Schwarzian action is just
\begin{equation}
  \label{eq:45}
  \partial_\timet\{y,\timet\}=0
\end{equation}
we see, as in the higher-dimensional cases mentioned in Section
\ref{sec:particle}, that the Schwarzian action just encodes mass
conservation.

In deriving the action \eqref{eq:10} we have not assumed anything about the
topology of the boundary. This is therefore the appropriate result if the
boundary is taken to be a line, i.e., the zero temperature result since the boundary
coordinate corresponds to Euclidean time. The finite temperature result is obtained
if the boundary is taken to be a circle with periodicity $\timet\sim \timet + \beta$.
Going once around the circle, the group element $g$ has to obey
\begin{equation}
  \label{eq:46}
  g(\timet+\beta)=g(\timet)
\end{equation}
in order to be single-valued. It is straightforward to see that this identification can be achieved by the field redefinition
\begin{equation}
  \label{eq:47}
  y=\tan\Big(\frac{\pi h(\timet)}{\beta}\Big)
\end{equation}
under which the Schwarzian action reads
\begin{equation}
  \label{eq:Schwarzfin}
  I[h]=\bar{X}\int \dd \timet\, \Big(\{h,\timet\}+\frac{2\pi^2}{\beta^2}(h')^2\Big)\,.
\end{equation}

From the point of view taken in this section,
all homogeneous spaces based on the algebra $\mathfrak{sl}(2,\R)$
can be equipped with the above boundary conditions that lead to a Schwarzian action on
the respective boundaries, modulo possible subtleties concerning the
topology of the bulk spacetime; see e.g., \cite{Cotler:2019nbi} for
the $\textrm{dS}_2$ case.\footnote{While the case of the
  two-dimensional light cone of Section~\ref{sec:bf-light-cone} has
  not been worked out in detail, it is highly suggestive in light of
  the results \cite{Carlip:2016lnw,Nguyen:2020hot}.}

\subsection{\texorpdfstring{AdS--Carroll$_2$}{AdS--Carroll2}}
\label{sec:AdSCaction}

We turn our attention to a non-Lorentzian spacetime. As apparent from
Table~\ref{tab:Liealg2}, the algebra of extended AdS--Carroll (AdSC) is isomorphic to the
centrally-extended 2d Poincar\'e algebra. The
interpretation of the generators, however, when viewed as a
homogeneous space is different. We expand the (co)adjoint vector
$\mathcal{X}$ and gauge field as
\begin{equation}
  \label{eq:AdSCdilaton}
\mathcal{X}=X^0\Ht+Z\Pt+X \Bt+X^1 \Mt \qquad A=e^{0}\Ht+e^{1}\Pt+\omega \Bt+Y \Mt
\end{equation}
where the components of the former are labelled such that $X^0$
enforces the torsion constraint for $e^0$ and so forth. In contrast to
the (Euclidean) $\textrm{AdS}_2$ case above, boundary condition for
this spacetime have not been discussed in the literature. We therefore
turn to the algebraic algorithm presented in Section
\ref{sec:constr-part-acti} in order to look for consistent boundary
conditions.

From the form of the algebra presented in Table~\ref{tab:Liealg2} it is unclear whether a two-dimensional non-abelian algebra exists, but after the basis change
\begin{equation}
    \label{eq:AdSCbasis}
    \Lt_{\pm}=\Ht\pm \Bt
\end{equation}
one finds two subalgebras spanned by $\{\Lt_+,\Pt\}$ and
$\{\Lt_-,\Pt\}$ respectively. We choose the latter as the subalgebra the generators of which are imposed as constraints (the former leads
again to identical conclusions upon redefinition of the radial
coordinate). The fixed element $K$ in \eqref{eq:39} is then required to be
of the form $K=\Lt_+ + \gamma_{\Mt}\,\Mt$, with $\gamma_{\Mt}$
arbitrary such that $\langle \Lt_-,K\rangle$ is non-zero and the constraints associated to $\Pt$ and $\Lt_-$ become second-class. This translates to boundary conditions reminiscent of highest-weight gauge
\begin{subequations}
    \label{eq:11}
\begin{align}
  A&=(e^{\rho}\Lt_{+}+\mathcal{L}(\timet)e^{-\rho}\Lt_{-}+\mathcal{T}(\timet)\Pt_1+ \gamma_{\Mt} \Mt)\dd \timet+\dd \rho \,\Pt_1\\
  \mathcal{X}&=e^\rho \,\bar{X}\, \Lt_++\mathcal{O}(1)
\end{align}
\end{subequations}
where the radial dependence is captured by the group element
$b=\exp (\rho \,\Pt)$, and the form of the dilaton field follows from
the universal boundary condition \eqref{eq:groupbc} where we set
$\partial_\timet f=\bar X$.

Under gauge transformations with generator
$\lambda=\lambda_+\Lt_++\lambda_-\Lt_-+\lambda_{\Pt} \Pt+\lambda_{\Mt}
\Mt$ that preserve the form of $A$ in \eqref{eq:11} the
state-dependent functions transform as
\begin{equation}
    \delta \mathcal{T}=\lambda_+ \mathcal{T}' + \lambda_+' \mathcal{T}-\lambda_+''\qquad \delta \mathcal{L}= \lambda_+ \mathcal{L}' + 2 \mathcal{L} \lambda_+'-\frac{1}{2}(\lambda_{\Mt}'\mathcal{T} + \lambda_{\Mt}'')\,.
\end{equation}
This transformation law is the hallmark of twisted warped symmetry.
Both central charge and $\mathfrak{u}(1)$ level are zero so
that the only non-trivial cocycle is the twist, $\kappa=1$ in the
conventions of \cite{Afshar:2015wjm,Afshar:2019tvp}. The fact that the
symmetry algebra $\mathfrak{iso}(1,1)^{c}$ is naturally related to
warped Virasoro symmetries was also found in the work
\cite{Hartong:2017bwq} that studied three-dimensional Chern--Simons
theory based on two copies of $\mathfrak{iso}(1,1)^{c}$ and uncovered a
warped Virasoro symmetry as asymptotic symmetry algebra. We therefore
expect the boundary action to be related to some version of the warped Schwarzian
action of~\cite{Afshar:2019tvp} at these specific values of the
central charges. We will find this expectation confirmed below.

In order to obtain some geometric insight into the boundary condition,
we can recover from \eqref{eq:11} the zweibein
\begin{equation}
  \label{eq:13}
  e^0=(e^{\rho}+\mathcal{M}e^{-\rho})\dd \timet\qquad\qquad e^1=\mathcal{T}\dd \timet+ \dd \rho
\end{equation}
and thus the Carrollian structure
\begin{equation}
  \label{eq:14}
  q=(\mathcal{T}\dd \timet+ \dd \rho\,)^2\qquad\qquad n=\frac{1}{e^{\rho}+\mathcal{M}e^{-\rho}}(\partial_\timet-\mathcal{T}\partial_\rho)\,
\end{equation}
where $q$ is the degenerate metric and $n$ denotes the vector field lying in its kernel.
This AdSC spacetime can be regarded as a null surface embedded in three-dimensional AdS space. More precisely, starting from Poincar\'e patch coordinates in three dimensions
\begin{equation}
  \label{eq:15}
  \dd s^2=\dd \rho^2- e^{\rho}\dd x^+\dd x^-
\end{equation}
the null-surface $x^-=0$ has induced degenerate metric and normal vector of the form
\begin{equation}
  \label{eq:16}
  q=\dd \rho^2\qquad\qquad n=e^{-\rho}\partial_+
\end{equation}
coinciding with \eqref{eq:14} in the case $\mathcal{T}=\mathcal{M}=0$ which one may regard as the vacuum configurations of \eqref{eq:11}. The boundary of the null-surface is obtained in the limit $\rho\rightarrow \infty$. The boundary coordinate $\timet$ of the BF theory is subsequently interpreted as a null-coordinate along the boundary of $\textrm{AdS}_3$. 

We address one more interesting feature of the AdSC geometry \eqref{eq:14}. When approaching the boundary $\rho \rightarrow \infty$, the degenerate metric $q$ diverges while the vector $n$ goes to zero. But upon introducing the boundary defining function $\Omega=e^{-\rho}$ and rescaling both quantities by the conformal factor $q\rightarrow \Omega^2q$, $n\rightarrow \Omega^{-1}n$, one finds that the pull-back of the AdSC structure to the boundary $\Omega=0$ yields
\begin{equation}
  \label{eq:AdSCboundary}
  q=0\qquad\qquad n=\partial_\timet\,.
\end{equation}
Since $\Omega$ is defined only up to a non-vanishing factor, one finds as the boundary
structure of AdSC spacetimes a conformal class of vectors together with the zero metric.
This is precisely the same boundary structure as 2d asymptotically flat spacetimes. So
not only share these geometries the same local symmetry group, i.e., the Poincar\'e group,
but also the same conformal boundary structure.

We turn now to the boundary action that can be obtained from the above
boundary conditions. Parametrizing the group element as
\begin{equation}
  \label{eq:17}
  g=e^{y \Lt_{+}}\,e^{\phi_-\Lt_{-}}\,e^{\phi_{\Pt} \Pt}\,e^{z\Zt}
\end{equation}
the current constraints implied by~\eqref{eq:11}
\begin{equation}
    g^{-1}\partial_\timet g|_+=1\qquad\qquad g^{-1}\partial_\timet g|_{\Mt}=\gamma_\Mt
\end{equation}
allow to solve $\phi_-, \phi_\Pt$ algebraically in terms of $y,z$ due to the second-class nature of the constraints.
Plugging the solutions into the action~\eqref{eq:9} leads to
\begin{equation}
  \label{eq:18}
  I=\bar{X}\int \dd \timet\, \left(z'(\log y')'-z''+\gamma_{\Mt}\,(\log y')'\right)\,.
\end{equation}
We recognize this as a form of the twisted warped action of \cite{Afshar:2019tvp,Afshar:2019axx}.\footnote{Upon redefining $y=e^{i u}$ this result agrees with the zero-temperature version of \cite{Gomis:2020wxp}.} In the parametrization \eqref{eq:17}, the two free functions in the connection \eqref{eq:11} are given by
\begin{equation}
  \label{eq:TandL}
  \mathcal{T}=-(\log y')'\qquad\qquad \mathcal{L}=\frac{1}{2}\left(z'(\log y')'-z''-\gamma_\Mt(\log y')'\right)
\end{equation}
and the equations of motion are equivalent to their conservation
\begin{equation}
  \label{eq:49}
  \partial_\timet \mathcal{T}=\partial_\timet \mathcal{L}=0\,.
\end{equation}
Finally, this action is invariant under the finite transformations
\begin{equation}
  \label{eq:28}
  y\mapsto a+y\, c\qquad\qquad z\mapsto z+d +b\,y\,
\end{equation}
that correspond to the global symmetry group of AdSC, i.e., $\textrm{ISO}(1,1)^{c}$. As in the Schwarzian case discussed above, these transformations should be interpreted as gauge transformation arising due to the redundancy in the reduction.

As a final comment we remind the reader that the AdSC theory discussed here exhibits the same algebra as dS-Galilei and 2d Minkowski space. Consequently, the boundary conditions \eqref{eq:11} are a priori also applicable to those theories albeit with different geometrical interpretation of generators and free fields. Indeed, the work \cite{Afshar:2019axx} found essentially the same action, but with complex fields, as boundary action in the case of (Euclidean) flat space and found an explicit relation to a scaling limit of the effective action of the complex SYK model.

\subsection{Other kinematical algebras}
\label{sec:remainingalg}

In the last two sections we constructed boundary actions for hyperbolic space/Euclidean $\textrm{AdS}_2$ and AdS--Carroll$_2$ as representatives for all kinematical spacetimes based on $\mathfrak{so}(2,1)$ and $\mathfrak{iso}(1,1)^{c}$. Disregarding the sphere and the static spacetime it remains to discuss spacetimes with symmetry algebras $\mathfrak{iso}(2)^{c}$ and $\textrm{Heisenberg}^{cc}$, i.e., spacetimes in the second and fourth block of Table~\ref{tab:Liealg2}, in order to cover all kinematical algebras with invariant metric.

According to the algorithm of Section \ref{sec:constr-part-acti} we should look for even-dimensional non-commutative subalgebras that we can use to write down boundary conditions that reduce the dynamics of the point particle action \eqref{eq:groupaction}. Starting with the doubly extended Heisenberg algebra that corresponds to e.g., flat Carroll spacetime, we find that the form of the algebra does not allow for any such subalgebra. The algebra is nilpotent from which follows that all two-dimensional subalgebras are abelian. Furthermore, the center of the algebra is two-dimensional so that any four-dimensional subalgebra necessarily contains an element that commutes with all of its remaining generators. Boundary conditions that derive from the algebraic structure and act by restricting the charges $Q$ \eqref{eq:32} will therefore lead to first-class constraints and gauge symmetry in the boundary action.

In the case of the algebra $\mathfrak{iso}(2)^{c}$ corresponding to e.g., AdS--Galilei, it is straightforward to show that no two-dimensional non-abelian subalgebra exists unless one allows the algebra, and thus the fields, to become complex.

This means that the procedure in Section \ref{sec:boundary} needs to be modified to construct boundary actions for these examples. The simplest `modification' is to not enforce any constraints at all, i.e., to stick with the most general boundary conditions. In this case the boundary action is always given by \eqref{eq:groupaction}.\footnote{Another option would be to not enforce the boundary condition \eqref{eq:groupbc}. The symplectic structure of any BF model, regardless of the existence of an invariant metric, can then be shown to lead on its boundary to the geometric action on the coadjoint orbit of the respective gauge group without any Hamiltonian; for more details see, e.g., Section 6.1 in \cite{Gonzalez:2018enk}.}

We leave the construction of alternative boundary actions for these cases for future work. 

\section{Applications and generalizations}
\label{sec:conclusions}

The main conclusion of our work is that it is done --- in particular, we have provided answers to the questions posed in the introduction: there is a NC version of 2d dilaton gravity, discussed in Section \ref{sec:NCandCdilaton}; the Schwarzian analogue for the AdS--Carroll$_2$ limit of JT gravity is the twisted warped action, discussed in Section \ref{sec:AdSCaction}.

Rather than summarizing our results in past tense, we address potential applications of various limits of JT gravity and their dilaton generalizations in Section \ref{sec:applications}, and point out some interesting and viable generalizations in Section \ref{sec:generalizations}. 

\subsection{Selected applications}\label{sec:applications}

Without claiming to be complete, here is a list of selected applications of our results, listed by order of appearance:
\begin{itemize}
    \item {\bf Light cone theory.} The light cone Lagrangian \eqref{eq:lc1} was easy enough to construct, but the physical interpretation of the latter remains to be explored. As advertised in~\cite{Matulich:2019cdo} the light cone theory might be interesting in relation to asymptotically flat spacetimes in three dimensions; cf.~\cite{Nguyen:2020hot} for the four-dimensional case. Thus, likely applications of \eqref{eq:lc1} are in the context of three-dimensional asymptotically flat gravity. Moreover, it could be interesting to study a dilaton gravity-inspired generalization of the light cone theory, analogous to \eqref{eq:jt1}.
    \item {\bf Newton--Cartan dilaton gravity.} The NC dilaton gravity action \eqref{eq:NCdilnew} has the same status for NC gravity as generic dilaton gravity \eqref{eq:jt1} (or its torsionful generalization below that equation) for Riemann--Cartan gravity. While we have shown numerous similarities --- the existence of constant and linear dilaton sectors, the exact solubility of the equations of motion, the possibility to accommodate higher-dimensional gravity models --- we have provided only one example. Given the vast literature of ordinary 2d dilaton gravity (see the table in \cite{Grumiller:2006rc} for a selected list of models) it seems likely that there are several interesting NC dilaton gravity models waiting to be applied.
    \item {\bf Carroll dilaton gravity.} We were slightly less explicit concerning Carroll dilaton gravity \eqref{eq:cardeal}, but clearly the same techniques that we used to study NC dilaton gravity can be applied there. Similar remarks concerning applications as in the previous item apply to this case, with the additional interesting option to explore the relationship to other field theories on null manifolds, see, e.g., \cite{Duval:2014uoa, Bagchi:2019clu}.
    \item {\bf The rest.} We were even less explicit regarding several other entries in Table~\ref{tab:Liealg2}, largely because we do not have a good proposal for applications of, say, the para-Galilei or the static case. Nevertheless, such applications may exist, and if they do, again the same techniques as for NC dilaton gravity can be applied to construct and solve JT-like and dilaton gravity-like models that build upon these homogeneous spaces.
    \item {\bf Boundary actions.} The loosest set of boundary conditions for any BF-type model always leads to a boundary action describing a particle on a group manifold \eqref{eq:groupaction}. Since for many gravity-inspired applications something like Brown--Henneaux boundary conditions is preferred, it will be rewarding to apply the Hamiltonian reduction of the boundary action described in Section \ref{sec:constr-part-acti} to other models of interest. Such applications will be analogous to the example of AdS--Carroll$_2$ we provided in Section \ref{sec:AdSCaction}, which led to a twisted warped action \eqref{eq:18}.
    \item {\bf Lifshitz, Schr\"odinger et al.} The theories mentioned in Appendices \ref{sec:lifshitz-schrodinger}, \ref{sec:coadjoint-theories} also lead to BF theories, and thus there will be `dilaton gravity' versions thereof, analogous to Section \ref{sec:NCandCdilaton}. Some of these models may be useful for applications in the context of Lifshitz or Schr\"odinger holography.
\end{itemize}

Besides the rather direct applications above there are also several exciting potential applications that will require --- in some cases substantial --- further input (here we order the items by likelihood of substantial progress):
\begin{itemize}
    \item {\bf JT/SYK-like correspondences.} Possibly the largest set of potential applications is to generalize the JT/SYK correspondence to interesting limiting cases. The fact that our models emerge from a limit of JT suggests that one can also implement a similar limit on the dual quantum mechanics model. A concrete example realizing this expectation is the flat space/cSYK correspondence discussed in \cite{Afshar:2019axx}. Of course, taking limits is just the first step in a much bigger picture. Many of the developments in the JT/SYK correspondence associated with chaos, relation to random matrix models, applications to quantum gravity, etc.~(see \cite{Mertens:2018fds,Sarosi:2017ykf,Gu:2019jub} for reviews) could be transposed to these limiting cases. We are convinced that this route can lead to numerous exciting discoveries in the near future.
    \item {\bf Relation to three-dimensional models.} There is a long-term relationship between three- and two-dimensional gravity, since the latter arises by dimensional reduction of the former. But the relation is deeper than that. Also at a technical level there are numerous similarities, particular in the respective gauge theoretic formulations as Chern--Simons and BF-theories --- both are topological gauge theories of Schwarz type and can feature non-trivial edge modes, depending on the boundary conditions. For metric BF theories the relation should be straightforward, since the operations of holographic reduction of three-dimensional Chern--Simons to 2d WZW at the boundary commutes with dimensional reduction and yields a corresponding holographic reduction of 2d BF to a corresponding boundary theory, see, e.g., \cite{Mertens:2018fds}. 
    Indeed many of the findings and tools used for $2+1$ dimensional Chern--Simons, like, e.g., in~\cite{Ozdemir:2019orp,Ravera:2019ize, Concha:2019lhn,Gomis:2019nih}, can be applied to BF theories.
    However, for WZW models without invariant metric the situation is not well understood in general; for example WZW models based on Lie algebras admit a Sugawara construction only when possessing an invariant metric~\cite{Mohammedi:1993rg,FigueroaO'Farrill:1994hx}. As r\^ole model we point here to a recent example by Chaturvedi, Papadimitriou, Song and Yu \cite{Chaturvedi:2020jyy} (see also refs.~therein), who discussed the dimensional reduction of AdS$_3$ Einstein gravity with Comp\'ere--Song--Strominger boundary conditions \cite{Compere:2013bya}, both on the gravity side and the SYK-side. Analogous applications and relations to higher-dimensional limiting theories should exist for the various limits of JT discussed in our work.
    \item {\bf Thermodynamics and Cardyology.} Theories without light cone may have a hard time of defining black hole-like entities. Nevertheless, some of these theories do feature highly entropic objects corresponding to finite temperature states, see e.g.~\cite{Bergshoeff:2016soe,Grumiller:2017sjh}, and regardless of their geometric interpretation it is of interest to understand their entropy and, whenever possible, provide a Cardy-inspired microstate counting. A first step in this direction could be a thermodynamical analysis starting with the free energy derived from the on-shell action, analogous to 2d dilaton gravity \cite{Grumiller:2007ju}. 
    \item {\bf Quantization and holography.} 
        Quantization of 2d NC dilaton gravity seems a feasible endeavor, due to the quantum integrability of BF-theories (see the review \cite{Birmingham:1991ty}), and might provide an interesting avenue to Galilean and Carrollian quantum gravity and holography. While this leads to various puzzles, such as  is there something like an information loss problem in any of these theories, we emphasize that quantizing all our models is possible and could lead to unexpected insights.
        \item{\bf Minkowski and AdS--Carroll.} As homogeneous spaces Minkowski and AdS--Carroll are both based on the Poincaré algebra and are connected via an exchange of boosts and spatial translations, a relationship that is true in any dimension~\cite{Figueroa-OFarrill:2018ilb}. The findings of Section~\ref{sec:AdSCaction} suggest that there might be a deeper relationship that remains to be explored. Two observations triggered our interest:
    1.) Since the cosmological constant, that gives AdS its `boxlike' properties, is still nonzero for AdS--Carroll it might inherit similar advantageous holographic features. 2.) Recent works on celestial amplitudes (for a review see~\cite{Strominger:2017zoo}) are based on boost eigenstates, rather than the more conventional momentum eigenstates. This therefore mirrors the above described exchange.
    \item {\bf Relation to $\boldsymbol{JT}$ deformations of JT?} The limits of JT gravity that we considered break/deform Lorentz symmetries and typically introduce an extra $\mathfrak{u}(1)$. Also Guica's $JT$ deformations \cite{Guica:2017lia} break/deform Lorentz symmetries and require an extra $\mathfrak{u}(1)$. Given that the more standard $T\bar T$-deformations upon dimensional reduction have a nice interpretation in terms of flow equations in 2d dilaton gravity \cite{Gross:2019ach,Gross:2019uxi,Grumiller:2020fbb}, we speculate that a similar relationship could hold between $JT$ deformations and certain limits of JT gravity. Addressing this last point would require a better understanding of the second and third item in this list. 
\end{itemize}

Most likely there are further potential applications that are missing in our lists above, but let us move on to our final point, generalizations.

\subsection{Selected generalizations}\label{sec:generalizations}

Here is a list of five classes of generalizations that we consider promising and viable:
\begin{itemize}
    \item {\bf More general boundary conditions.} The algorithm explained in Section \ref{sec:constr-part-acti} imposed a number of assumptions. In particular, $K$ and $\gamma$ in the constraint \eqref{eq:39} were assumed to be state-independent, and the constraint algebra \eqref{eq:40} was enforced to be second class. Neither of these assumptions is mandatory, so by relaxing either of them it is possible to construct a whole menagerie of new boundary conditions. A full classification of all possibilities could be worthwhile, for novel applications and for purely theoretical reasons, to get a better understanding of the landscape of boundary conditions.  
    \item {\bf Additional topological fields.} Our focus was on gravity variables, zweibein and connection, but for some applications it can be of interest to add non-abelian gauge fields. Such a generalization is straightforward in the BF-formulation and will lead to new conserved charges and additional boundary degrees of freedom, but will maintain the topological nature of the theory.
    \item {\bf Supersymmetry and/or higher spins.} A variant of the previous item is to include supersymmetry and consider various limits of 2d supergravity theories and their boundary actions (see \cite{Cardenas:2018krd} and refs.~therein). Similarly, one can extend to higher spin gravity (along the lines of \cite{Gonzalez:2018enk,Alkalaev:2020kut}) and take various limits analogous to the present work. In all these generalizations one would still keep the topological nature of the underlying theory.
    \item {\bf Non-linear gauge symmetries.} Chern--Simons models in 3d are rigid, i.e., their most general consistent deformation (in the sense of Barnich and Henneaux \cite{Barnich:1993vg}) is another Chern--Simons model with the same number of gauge symmetries. A similar story applies to 2d BF theories, but with an interesting extension: the most general consistent deformation of 2d BF is a non-linear gauge theory \cite{Ikeda:1993fh} known as Poisson-sigma model (PSM) \cite{Schaller:1994es}. The Poisson-sigma model Lagrangian
    \begin{equation}
        {\mathcal L}_{\textrm{\tiny{PSM}}}[X^I,\,A_I] = X^I\dd A_I + \frac12\,P^{IJ}(X^K)\,A_I\wedge A_J
    \end{equation}
    features again scalars $X^I$ and connection 1-forms $A_I$ and is first order in derivatives, just like BF theories, but the interpretation of the $X^I$ is now as target space coordinates spanning a Poisson manifold. The main new ingredient is an arbitrary Poisson-tensor $P^{IJ}=-P^{JI}$, subject to the non-linear Jacobi identities $P^{IL}\partial_L P^{JK}+\textrm{cycl}(I,J,K)=0$. In this languages, BF theories (with or without metric) are merely special cases of PSMs with linear Poisson tensor, and the limits we have taken can be rephrased as corresponding limits of the Poisson tensor. Secretly, we have already introduced specific PSMs when providing the action for 2d dilaton gravity \eqref{eq:jt1}, NC dilaton gravity \eqref{eq:NCdilnew} [see particularly the non-linear gauge symmetries \eqref{eq:nonlin}] and Carroll dilaton gravity \eqref{eq:cardeal}. In order to construct boundary actions for all these models it will be necessary to understand how this can be done in generic PSMs, which is why this particular generalization seems eminently worthwhile.
    \item {\bf Adding matter.} Topological gauge theories like BF or PSM have numerous technical advantages, but miss an important aspect of physics, namely locally propagating physical degrees of freedom. A possible compromise between the conflicting desires of keeping the model simple and physically rich is to couple BF theories (or PSMs) to matter degrees of freedom. Some consequences of such a coupling are reviewed in the context of 2d dilaton gravity in \cite{Grumiller:2002nm}. While ambitious, it could pay off to add matter to theories like NC or Carroll dilaton gravity and address some questions concerning 1-loop effects, backreactions, etc. 
\end{itemize}

While further generalizations not envisaged here are conceivable, the lists above provide already a plethora of possibilities for future research.

\subsection*{Acknowledgments}
\label{sec:acknoledgment}

We thank Arash Ranjbar, Dieter Van den Bleeken and Jos\'e
Figueroa-O'Farrill for useful discussions. We want to thank Jos\'e
Figueroa-O'Farrill for valuable advice concerning metric Lie algebras.

DG was supported by the Austrian Science Fund (FWF), projects
P~30822-N27 and P~32581-N27. During the start of this project the
research of SP was supported by the ERC Advanced Grant
``High-Spin-Grav'' and by FNRS-Belgium (Convention FRFC PDR T.1025.14
and Convention IISN 4.4503.15). SP was supported by the Leverhulme Trust
Research Project Grant (RPG-2019-218) ``What is Non-Relativistic
Quantum Gravity and is it Holographic?''. The work of JH is supported
by the Royal Society University Research Fellowship ``Non-Lorentzian
Geometry in Holography'' (grant number UF160197). JS was supported by
the Erwin-Schr\"odinger fellowship J-4135 of the Austrian Science Fund
(FWF) and by the NSF grant 1707938.

\appendix

\section{Metric Lie algebras of low dimension}
\label{sec:metric-lie}

We summarize low-dimensional real Lie algebras that admit an invariant
metric (see, e.g., \cite{Ovando:Review} for a review). The Roman
numerals refer to the name of specific Lie algebras according to
Bianchi's classification~\cite{Bianchi,MR1900159}, c.f., our summary
Table~\ref{tab:Liealg2}.

\subsection*{\texorpdfstring{$\boldsymbol{\dim (\mathfrak{g})=1}$}{dim(g)=1} and \texorpdfstring{$\boldsymbol{\dim (\mathfrak{g})=2}$}{dim(g)=2}}

For one and two dimensional Lie algebras solely the abelian Lie
algebras admit an invariant metric.

\subsection*{\texorpdfstring{$\boldsymbol{\dim (\mathfrak{g})=3}$}{dim(g)=3}}

Three dimensional Lie algebras that admit an invariant metric are
either abelian (denoted by Bianchi as $\text{I}$) or simple and amount
to three distinct cases in total. The simple one is either
$\mathfrak{sl}(2,\R) \simeq \mathfrak{so}(2,1) \simeq
\mathfrak{so}(1,2) \simeq \mathfrak{su}(1,1) \simeq \text{VII}$, which
correspond to the (A)dS and light cone cases or
$\mathfrak{so}(3)\simeq \mathfrak{su}(2) \simeq \text{IX}$ leading to
the sphere (or Euclidean de Sitter), see Table~\ref{tab:Liealg2}.

\subsection*{\texorpdfstring{$\boldsymbol{\dim (\mathfrak{g})=4}$}{dim(g)=4}}

There exist five metric Lie algebras of dimension four. One of them is
the abelian Lie algebra. Additionally, there are the trivial central
extensions of the two simple three dimensional cases
$\R \times \mathfrak{sl}(2,\R)$ and $\R \times \mathfrak{so}(3,\R)$.
They serve in this work as a starting point for contractions to the
remaining two metric solvable algebras given by the (centrally)
extended Euclidean $\mathfrak{iso}(2)^{c}=\text{VII}_{0}^{c}$ or
Poincar\'e algebra $\mathfrak{iso}(1,1)^{c}=\text{VI}_{0}^{c}$ (which
are isomorphic to their (A)dS-Carrollian and (A)dS-Galilean cousins,
c.f., Table~\ref{tab:Liealg2}). The centrally extended Poincar\'e
algebra is also known as Maxwell or Boidol algebra and the Euclidean
case is sometimes referred to as oscillator algebra.

\subsection*{\texorpdfstring{$\boldsymbol{\dim(\mathfrak{g})=5}$}{dim(g)=5}}

For five dimensional Lie algebras there are six Lie algebras that
admit an invariant metric, five of which are trivial central
extensions of the four dimensional ones. The remaining unique
indecomposable metric Lie algebra is
$\mathrm{Heisenberg}^{cc}\simeq \mathrm{II}^{cc}$ of our
Table~\ref{tab:Liealg2} and underlies the doubly extended Carrollian
and Galilean theories.

\subsection*{\texorpdfstring{$\boldsymbol{\dim(\mathfrak{g})>5}$}{dim(g)>5}}

Beyond dimension five one can use the classifications of metric Lie
algebras that have been obtained for signature $(n,1)$~\cite{Medina}
(see also Section 4 in \cite{Figueroa-OFarrill:2003fkz}), $(n,2)$
\cite{KathOlb:2004} and $(n,3)$~\cite{KathOlb:2006}. Kinematical Lie
algebras and their relation to metric Lie algebras in any dimension
have also been studied in~\cite{Matulich:2019cdo}. The
higher-dimensional algebras do not play any r\^ole in the present
work.

\section{Lifshitz, Schr\"odinger and \texorpdfstring{$\boldsymbol{1/c}$}{1/c} expanded BF theories}
\label{sec:lifshitz-schrodinger}

Here we provide all the necessary data to construct Lifshitz,
Schr\"odinger and $1/c$ expanded BF theories. None of them (even with
central extensions taken into account) lead to metric BF theories.

The Lifshitz algebra is spanned by dilatations ($\Dt$), time ($\Ht$)
and space ($\Pt$) translations
\begin{subequations}
  \label{eq:lif}
  \begin{align}
  [\Dt,\Ht] &= z \Ht &
  [\Dt,\Pt] &= \Pt 
  \end{align}
\end{subequations}
with the most general (degenerate) invariant bilinear form given by
\begin{align}
  \label{eq:lifmetr}
 \langle \Dt,\Dt \rangle&=\chi \, .
\end{align}
This algebra has no nontrivial central extensions.

For $z=2$ this algebra can be enhanced to the Schr\"odinger algebra by
adding mass $\Mt$ and special conformal transformations $\Ct$,
\begin{subequations}
 \label{eq:schr}
 \begin{align}
   [\Bt,\Pt] &= \Mt  &  [\Dt,\Ht] &= 2 \,\Ht\\
   [\Dt,\Pt] &= \Pt &  [\Dt,\Bt] &= -\Bt \\
   [\Ct,\Pt] &= \Bt & [\Ct,\Dt] &= 2 \, \Ct  \, .
 \end{align}
\end{subequations}
This algebra has the same bilinear form as the Lifshitz case, but
admits one central extension $[\Ht,\Ct]=\Zt$. Even including this
central extension leaves the bilinear form degenerate.

The nonzero commutation relations of the $1/c$ expanded Poincar\'e algebra (till level $1$) are given by~\cite{Hansen:2018ofj}
\begin{align}
  \label{eq:1cexp}
  [\Bt,\Ht] &= \Pt &
  [\Bt,\Pt] &= \Ht^{(1)} &
  [\Bt,\Ht^{(1)}] &= \Pt^{(1)} &
  [\Bt^{(1)},\Ht] &= \Pt^{(1)}
\end{align}
with the most general degenerate invariant bilinear form
\begin{subequations}
\begin{align}
  \label{eq:1cmetr}
  \langle \Bt,\Bt \rangle&=\chi_{\Bt} &
  \langle \Bt,\Ht \rangle&=\chi_{\Bt\Ht} &
  \langle \Ht,\Ht \rangle&=\chi_{\Ht} & \\
  \langle \Bt,\Bt^{(1)} \rangle&=\chi_{\Bt \Bt^{(1)}} &
  \langle \Ht,\Bt^{(1)} \rangle&=\chi_{\Ht \Bt^{(1)}} &
  \langle \Bt^{(1)},\Bt^{(1)} \rangle&=\chi_{\Bt^{(1)}} \, .
\end{align}
\end{subequations}
These algebras appear in the context of Lie algebra expansions (see,
e.g., \cite{deAzcarraga:2002xi,Izaurieta:2006zz,Khasanov:2011jr})
which have recently found applications in the context of general
relativity. The notation can be explained as follows. Take the Cartan
connection 1-form, say
\begin{equation}
   A= \Bt \omega+\tau \Ht+e \Pt +\cdots
\end{equation}
where the dots denote possibly other elements in the Lie algebra. When
the 1-forms $\tau$, $e$ etc.\ depend on the speed of light (as is
the case for the Poincar\'e algebra) we can assume that the $c$
dependence is analytic so that we can Taylor expand
\begin{equation}
    \tau=\sum_{n=0}^\infty \tau_{(n)}c^{-2n}
\end{equation}
where we assume only even powers of $c^{-1}$. Substituting this
expansion into the Cartan connection and defining a Lie algebra
generator for each $\tau_{(n)}$ we end up with the Lie algebra
elements $\Ht^{(n)}=\Ht c^{-2n}$ (and analog for the remaining
generators). Since the structure constants can also depend on $c^{-2}$
it is possible that the bracket of a level $n$ with a level $m$
generator gives a generator of a level that is strictly larger than
$m+n$. The terms with superscript $(1)$ denote generators of next to
leading order (`level 1'). The level $0$ algebra, where one
quotients out all Lie algebra elements of level bigger than $0$ is
again the Galilei algebra. The algebra \eqref{eq:1cexp} is the algebra
where all levels $n>1$ have been modded out.

The $1/c$ expanded algebra in $1+1$ dimensions allows for $6$
nontrivial central extensions $\Mt$, $\Zt_{1}$ to $\Zt_{5}$
\begin{subequations}
\begin{align}
  [\Bt,\Ht] &= \Pt &
  [\Bt,\Pt] &= \Ht^{(1)} &
  [\Bt,\Ht^{(1)}] &= \Pt^{(1)} + \Zt_{1} &
  [\Bt^{(1)},\Ht] &= \Pt^{(1)} + \Zt_{2}\\
  [\Ht,\Pt] &= \Mt &
  [\Ht,\Pt^{(1)}] &= \Zt_{3} &
                               [\Pt,\Ht^{(1)}] &= -\Zt_{3}  &
  [\Bt,\Bt^{(1)}] &= \Zt_{4}  \\
&&  [\Bt,\Pt^{(1)}] &= \Zt_{6} &
                               [\Pt,\Bt^{(1)}] &= -\Zt_{6}  \, .
\end{align}
\end{subequations}
A necessary condition for a non-degenerate invariant metric would be to
only add two central extensions to \eqref{eq:1cexp}, c.f.,
\eqref{eq:neccriterion}. We have checked that the addition of $\Mt$
and any of the other central extensions does not lead to such a
non-degenerate invariant metric.

\section{Coadjoint theories and their limits}
\label{sec:coadjoint-theories}

For completeness we briefly mention another class of theories
where the invariant metric stays, basically by construction,
non-degenerate under limits. These theories have the distinctive
feature that the algebraic structure generalizes to generic dimension
(see, e.g.,~\cite{Matulich:2019cdo}).

They are based on so called coadjoint Lie algebras, which are a
subcase of the already mentioned double extension~\cite{Medina1985}.
Given a Lie algebra $\mathfrak{g}$ they are defined on the vector
space direct sum $\mathfrak{g} \dot +\mathfrak{g}^{*}$ (spanned by
$\ett_{\alpha}$ and $\ett^{*\alpha}$, respectively) by the commutation
relations
\begin{align}
 [\ett_{\alpha},\ett_{\beta}]                      & =c\indices{_{\alpha \beta}^{\gamma}} \ett_{\gamma}
                                                 & 
 [\ett_{\alpha},\ett^{*\beta}]                     & =-c\indices{_{\alpha \gamma}^{\beta}} \ett^{*\gamma}
                                                 & 
  [\ett^{*\alpha},\ett^{*\beta}]                   & =0
\end{align}
and the invariant metric
\begin{align}
  \label{eq:metriccoadjoint}
  \langle \ett_{\alpha}, \ett_{\beta} \rangle &= h_{\alpha\beta} &
  \langle \ett_{\alpha}, \ett^{* \beta} \rangle &= \delta_{\alpha}^{\beta}
\end{align}
which is non-degenerate due to the second term. The other part of the
invariant metric $h_{\alpha\beta}$ is some arbitrary (possibly
degenerate or zero) invariant symmetric bilinear form on
$\mathfrak{g}$. The extension of $\mathfrak{g}$
by $\mathfrak{g}^{*}$ is not central in general.

We discuss all limits and theories at once. Starting with (A)dS this
means that additionally to commutation relations \eqref{eq:adsstart}
we have the nonzero commutators
\begin{subequations}
  \label{eq:coadj}
\begin{align}
  [\Bt,\Ht^{*}] &= - \cc^{2}\Pt^{*} & [\Bt, \Pt^{*}] &= \pm \tc^{2} \Ht^{*} \\
  [\Ht,\Bt^{*}] &=  \lc \Pt^{*} & [\Ht, \Pt^{*}] &= \mp \tc^{2} \Bt^{*} \\
  [\Pt,\Bt^{*}] &= - \lc\Ht^{*} & [\Pt, \Ht^{*}] &= \cc^{2} \Bt^{*}
\end{align}
\end{subequations}
and in addition to the invariant metric~\eqref{eq:adsmetrst}, which
corresponds to the $h_{\alpha\beta}$ part described above, we have by
construction
\begin{align}
  \label{eq:coadjinv}
\langle \Bt,\Bt^{*} \rangle  &= 1 &\langle \Ht,\Ht^{*} \rangle  &= 1 &\langle \Pt,\Pt^{*} \rangle  &= 1 \, .
\end{align}
As already discussed, the part of the invariant metric given
by~\eqref{eq:adsmetrst} is degenerate under limits; it
is~\eqref{eq:coadjinv} that guarantees the existence of the
non-degenerate invariant metric. The limit of the algebra
$\mathfrak{g}$ induces the limits of $\mathfrak{g}^{*}$, which are
well defined by construction (a fact that
generalizes under certain circumstances to double
extensions~\cite{Prohazka:2017pkc}).

Having provided the Lie algebra and the invariant metric it is now an
easy exercise to write down the action by just inserting
into the Lagrangian~\eqref{eq:mBF}. There are two curious features of the BF
theories based on coadjoint Lie algebras. First, the existence of the
algebras and their invariant metric is not constrained to $1+1$
dimensions and as such these are the algebras that are interesting
candidates for generalization to higher dimension (see Section 7
of~\cite{Matulich:2019cdo} for details). Second, a comment in
relation to $2+1$ dimensions and Chern--Simons theories: The
coadjoint Lie algebras are precisely the ones that can be written as
$2+1$ dimensional BF theories, see, e.g., Section
6.2.2.~\cite{Birmingham:1991ty}. In particular, the three-dimensional Poincar\'e algebra can be regarded as the coadjoint algebra of $\mathfrak{so}(3)$ so that its Chern--Simons theory can be equivalently regarded as a (2+1)-dimensional BF theory of $\mathfrak{so}(3)$. For further recent works based on
these algebras we refer
to~\cite{Barducci:2019jhj,Bergshoeff:2020fiz,Barducci:2020blv}.

\section{Matrix representations}

We collect in this appendix matrix representations that are useful in
the calculations of Section \ref{sec:boundary}. In the following, let
$X^{(n)}_{ij}$ denote the $n\times n$ matrix with entry
$1$ in the slot $ij$ and zero everywhere else.

The $\mathfrak{sl}(2,\mathbb{R})$ basis $\Lt_+,\Lt_-,\Lt_0$ with
commutation relations
\begin{equation}\label{eq:sl2r}
    [\Lt_m,\Lt_n]=(m-n)\Lt_{m+n}\qquad m,n=-1,0,1
\end{equation}
used in the calculation of Section \ref{sec:schwarzianAdS} is
conveniently represented as
\begin{equation}
  \Lt_+= X^{(2)}_{21}
  \qquad
  \Lt_-= -X^{(2)}_{12}
  \qquad
  \Lt_0=\frac{1}{2} (X^{(2)}_{11}- X^{(2)}_{22})
\end{equation}
The invariant bilinear in this basis is given by
\begin{equation}
    \langle \Lt_+,\Lt_-\rangle=2\qquad\qquad \langle \Lt_0,\Lt_0\rangle=-1\,.
\end{equation}

The AdSC algebra in the basis $\Lt_+,\Lt_-,\Pt,\Mt$ used in Section
\ref{sec:AdSCaction} with commutation relations
\begin{equation}
  [\Lt_\pm,\Pt]=\pm\Lt_\pm\qquad\qquad
  [\Lt_+,\Lt_-]=-2\Mt
\end{equation}
can be represented by
\begin{equation}
  \Lt_+=X^{(3)}_{12}\qquad
  \Lt_-=X^{(3)}_{23}\qquad
  \Pt=X^{(3)}_{22}\qquad
  \Mt_+=-2X^{(3)}_{13}\,.
\end{equation}
The invariant bilinear form in this basis is given by
\begin{equation}
      \langle \Lt_+,\Lt_-\rangle=2\qquad\qquad \langle \Pt,\Mt\rangle=1\,.
\end{equation}

Finally, the doubly extended Heisenberg algebra
has the matrix representation
\begin{equation}
  \Bt= X^{(5)}_{12}-X^{(5)}_{35}\,\qquad
  \Pt=X^{(5)}_{13}+X^{(5)}_{25}\,\qquad
  \Zt=-X^{(5)}_{14}\,\qquad
  \Mt=2 X^{(5)}_{15}\,\qquad
  \Ht=X^{(5)}_{23}+X^{(5)}_{34}\,.
\end{equation}

\bibliographystyle{utphys} 
\bibliography{bibl} 

\providecommand{\href}[2]{#2}\begingroup\raggedright\begin{thebibliography}{100}

\bibitem{Barbashov:1980bm}
B.~Barbashov, V.~Nesterenko, and A.~Chervyakov, ``The solitons in some
  geometrical field theories,''
  \href{http://dx.doi.org/10.1007/BF01019238}{{\em Theor. Math. Phys.}
  {\bfseries 40} (1979) 572--581}.

\bibitem{Dhoker:1982wmk}
E.~D'Hoker and R.~Jackiw, ``{Liouville Field Theory},''
  \href{http://dx.doi.org/10.1103/PhysRevD.26.3517}{{\em Phys. Rev. D}
  {\bfseries 26} (1982) 3517}.

\bibitem{Teitelboim:1983ux}
C.~Teitelboim, ``{Gravitation and Hamiltonian Structure in Two Space-Time
  Dimensions},'' \href{http://dx.doi.org/10.1016/0370-2693(83)90012-6}{{\em
  Phys. Lett. B} {\bfseries 126} (1983) 41--45}.

\bibitem{Jackiw:1984je}
R.~Jackiw, ``{Lower Dimensional Gravity},''
  \href{http://dx.doi.org/10.1016/0550-3213(85)90448-1}{{\em Nucl. Phys. B}
  {\bfseries 252} (1985) 343--356}.

\bibitem{Mann:1989gh}
R.~B. Mann, A.~Shiekh, and L.~Tarasov, ``{Classical and Quantum Properties of
  Two-dimensional Black Holes},''
  \href{http://dx.doi.org/10.1016/0550-3213(90)90265-F}{{\em Nucl. Phys. B}
  {\bfseries 341} (1990) 134--154}.

\bibitem{Cadoni:1998sg}
M.~Cadoni and S.~Mignemi, ``{Entropy of 2-D black holes from counting
  microstates},'' \href{http://dx.doi.org/10.1103/PhysRevD.59.081501}{{\em
  Phys. Rev. D} {\bfseries 59} (1999) 081501},
  \href{http://arxiv.org/abs/hep-th/9810251}{{\ttfamily arXiv:hep-th/9810251}}.

\bibitem{Cacciatori:2002ib}
S.~Cacciatori, A.~Chamseddine, D.~Klemm, L.~Martucci, W.~Sabra, and D.~Zanon,
  ``{Noncommutative gravity in two dimensions},''
  \href{http://dx.doi.org/10.1088/0264-9381/19/15/310}{{\em Class. Quant.
  Grav.} {\bfseries 19} (2002) 4029--4042},
  \href{http://arxiv.org/abs/hep-th/0203038}{{\ttfamily arXiv:hep-th/0203038}}.

\bibitem{Grumiller:2007ju}
D.~Grumiller and R.~McNees, ``{Thermodynamics of black holes in two (and
  higher) dimensions},''
  \href{http://dx.doi.org/10.1088/1126-6708/2007/04/074}{{\em JHEP} {\bfseries
  04} (2007) 074}, \href{http://arxiv.org/abs/hep-th/0703230}{{\ttfamily
  arXiv:hep-th/0703230}}.

\bibitem{Sen:2008yk}
A.~Sen, ``{Entropy Function and AdS(2) / CFT(1) Correspondence},''
  \href{http://dx.doi.org/10.1088/1126-6708/2008/11/075}{{\em JHEP} {\bfseries
  11} (2008) 075}, \href{http://arxiv.org/abs/0805.0095}{{\ttfamily
  arXiv:0805.0095 [hep-th]}}.

\bibitem{Hartman:2008dq}
T.~Hartman and A.~Strominger, ``{Central Charge for AdS(2) Quantum Gravity},''
  \href{http://dx.doi.org/10.1088/1126-6708/2009/04/026}{{\em JHEP} {\bfseries
  04} (2009) 026}, \href{http://arxiv.org/abs/0803.3621}{{\ttfamily
  arXiv:0803.3621 [hep-th]}}.

\bibitem{Castro:2008ms}
A.~Castro, D.~Grumiller, F.~Larsen, and R.~McNees, ``{Holographic Description
  of AdS(2) Black Holes},''
  \href{http://dx.doi.org/10.1088/1126-6708/2008/11/052}{{\em JHEP} {\bfseries
  11} (2008) 052}, \href{http://arxiv.org/abs/0809.4264}{{\ttfamily
  arXiv:0809.4264 [hep-th]}}.

\bibitem{Maldacena:2016upp}
J.~Maldacena, D.~Stanford, and Z.~Yang, ``{Conformal symmetry and its breaking
  in two dimensional Nearly Anti-de-Sitter space},''
  \href{http://dx.doi.org/10.1093/ptep/ptw124}{{\em PTEP} {\bfseries 2016}
  no.~12, (2016) 12C104}, \href{http://arxiv.org/abs/1606.01857}{{\ttfamily
  arXiv:1606.01857 [hep-th]}}.

\bibitem{Mertens:2018fds}
T.~G. Mertens, ``{The Schwarzian theory — origins},''
  \href{http://dx.doi.org/10.1007/JHEP05(2018)036}{{\em JHEP} {\bfseries 05}
  (2018) 036},
\href{http://arxiv.org/abs/1801.09605}{{\ttfamily arXiv:1801.09605 [hep-th]}}.

\bibitem{Sarosi:2017ykf}
G.~Sárosi, ``{AdS$_{2}$ holography and the SYK model},''
  \href{http://dx.doi.org/10.22323/1.323.0001}{{\em PoS} {\bfseries Modave2017}
  (2018) 001},
\href{http://arxiv.org/abs/1711.08482}{{\ttfamily arXiv:1711.08482 [hep-th]}}.

\bibitem{Gu:2019jub}
Y.~Gu, A.~Kitaev, S.~Sachdev, and G.~Tarnopolsky, ``{Notes on the complex
  Sachdev-Ye-Kitaev model},''
  \href{http://dx.doi.org/10.1007/JHEP02(2020)157}{{\em JHEP} {\bfseries 02}
  (2020) 157}, \href{http://arxiv.org/abs/1910.14099}{{\ttfamily
  arXiv:1910.14099 [hep-th]}}.

\bibitem{Cotler:2016fpe}
J.~S. Cotler, G.~Gur-Ari, M.~Hanada, J.~Polchinski, P.~Saad, S.~H. Shenker,
  D.~Stanford, A.~Streicher, and M.~Tezuka, ``{Black Holes and Random
  Matrices},'' \href{http://dx.doi.org/10.1007/JHEP05(2017)118}{{\em JHEP}
  {\bfseries 05} (2017) 118}, \href{http://arxiv.org/abs/1611.04650}{{\ttfamily
  arXiv:1611.04650 [hep-th]}}. [Erratum: JHEP 09, 002 (2018)].

\bibitem{Saad:2019lba}
P.~Saad, S.~H. Shenker, and D.~Stanford, ``{JT gravity as a matrix integral},''
  \href{http://arxiv.org/abs/1903.11115}{{\ttfamily arXiv:1903.11115
  [hep-th]}}.

\bibitem{Dubovsky:2017cnj}
S.~Dubovsky, V.~Gorbenko, and M.~Mirbabayi, ``{Asymptotic fragility, near
  AdS$_{2}$ holography and $ T\overline{T} $},''
  \href{http://dx.doi.org/10.1007/JHEP09(2017)136}{{\em JHEP} {\bfseries 09}
  (2017) 136},
\href{http://arxiv.org/abs/1706.06604}{{\ttfamily arXiv:1706.06604 [hep-th]}}.

\bibitem{Cardy:2018sdv}
J.~Cardy, ``{The $ T\overline{T} $ deformation of quantum field theory as
  random geometry},'' \href{http://dx.doi.org/10.1007/JHEP10(2018)186}{{\em
  JHEP} {\bfseries 10} (2018) 186},
  \href{http://arxiv.org/abs/1801.06895}{{\ttfamily arXiv:1801.06895
  [hep-th]}}.

\bibitem{Maldacena:2018lmt}
J.~Maldacena and X.-L. Qi, ``{Eternal traversable wormhole},''
  \href{http://arxiv.org/abs/1804.00491}{{\ttfamily arXiv:1804.00491
  [hep-th]}}.

\bibitem{Goto:2018iay}
K.~Goto, H.~Marrochio, R.~C. Myers, L.~Queimada, and B.~Yoshida, ``{Holographic
  Complexity Equals Which Action?},''
  \href{http://dx.doi.org/10.1007/JHEP02(2019)160}{{\em JHEP} {\bfseries 02}
  (2019) 160}, \href{http://arxiv.org/abs/1901.00014}{{\ttfamily
  arXiv:1901.00014 [hep-th]}}.

\bibitem{Harlow:2018tqv}
D.~Harlow and D.~Jafferis, ``{The Factorization Problem in Jackiw-Teitelboim
  Gravity},'' \href{http://dx.doi.org/10.1007/JHEP02(2020)177}{{\em JHEP}
  {\bfseries 02} (2020) 177}, \href{http://arxiv.org/abs/1804.01081}{{\ttfamily
  arXiv:1804.01081 [hep-th]}}.

\bibitem{Kitaev:2018wpr}
A.~Kitaev and S.~J. Suh, ``{Statistical mechanics of a two-dimensional black
  hole},'' \href{http://dx.doi.org/10.1007/JHEP05(2019)198}{{\em JHEP}
  {\bfseries 05} (2019) 198}, \href{http://arxiv.org/abs/1808.07032}{{\ttfamily
  arXiv:1808.07032 [hep-th]}}.

\bibitem{Penington:2019npb}
G.~Penington, ``{Entanglement Wedge Reconstruction and the Information
  Paradox},'' \href{http://dx.doi.org/10.1007/JHEP09(2020)002}{{\em JHEP}
  {\bfseries 09} (2020) 002}, \href{http://arxiv.org/abs/1905.08255}{{\ttfamily
  arXiv:1905.08255 [hep-th]}}.

\bibitem{Almheiri:2019psf}
A.~Almheiri, N.~Engelhardt, D.~Marolf, and H.~Maxfield, ``{The entropy of bulk
  quantum fields and the entanglement wedge of an evaporating black hole},''
\href{http://arxiv.org/abs/1905.08762}{{\ttfamily arXiv:1905.08762 [hep-th]}}.

\bibitem{Almheiri:2019hni}
A.~Almheiri, R.~Mahajan, J.~Maldacena, and Y.~Zhao, ``{The Page curve of
  Hawking radiation from semiclassical geometry},''
  \href{http://dx.doi.org/10.1007/JHEP03(2020)149}{{\em JHEP} {\bfseries 03}
  (2020) 149}, \href{http://arxiv.org/abs/1908.10996}{{\ttfamily
  arXiv:1908.10996 [hep-th]}}.

\bibitem{Almheiri:2019yqk}
A.~Almheiri, R.~Mahajan, and J.~Maldacena, ``{Islands outside the horizon},''
  \href{http://arxiv.org/abs/1910.11077}{{\ttfamily arXiv:1910.11077
  [hep-th]}}.

\bibitem{Almheiri:2019qdq}
A.~Almheiri, T.~Hartman, J.~Maldacena, E.~Shaghoulian, and A.~Tajdini,
  ``{Replica Wormholes and the Entropy of Hawking Radiation},''
  \href{http://dx.doi.org/10.1007/JHEP05(2020)013}{{\em JHEP} {\bfseries 05}
  (2020) 013}, \href{http://arxiv.org/abs/1911.12333}{{\ttfamily
  arXiv:1911.12333 [hep-th]}}.

\bibitem{Penington:2019kki}
G.~Penington, S.~H. Shenker, D.~Stanford, and Z.~Yang, ``{Replica wormholes and
  the black hole interior},'' \href{http://arxiv.org/abs/1911.11977}{{\ttfamily
  arXiv:1911.11977 [hep-th]}}.

\bibitem{Grumiller:2002nm}
D.~Grumiller, W.~Kummer, and D.~Vassilevich, ``{Dilaton gravity in
  two-dimensions},''
  \href{http://dx.doi.org/10.1016/S0370-1573(02)00267-3}{{\em Phys. Rept.}
  {\bfseries 369} (2002) 327--430},
  \href{http://arxiv.org/abs/hep-th/0204253}{{\ttfamily arXiv:hep-th/0204253}}.

\bibitem{Callan:1992rs}
C.~G. Callan, Jr., S.~B. Giddings, J.~A. Harvey, and A.~Strominger,
  ``{Evanescent black holes},''
  \href{http://dx.doi.org/10.1103/PhysRevD.45.R1005}{{\em Phys. Rev.}
  {\bfseries D45} no.~4, (1992) R1005},
\href{http://arxiv.org/abs/hep-th/9111056}{{\ttfamily arXiv:hep-th/9111056
  [hep-th]}}.

\bibitem{Achucarro:1987vz}
A.~Achucarro and P.~Townsend, ``{A Chern-Simons Action for Three-Dimensional
  anti-De Sitter Supergravity Theories},''
\href{http://dx.doi.org/10.1016/0370-2693(86)90140-1}{{\em Phys.Lett.}
  {\bfseries B180} (1986) 89}.

\bibitem{Witten:1988hc}
E.~Witten, ``{(2+1)-Dimensional Gravity as an Exactly Soluble System},''
\href{http://dx.doi.org/10.1016/0550-3213(88)90143-5}{{\em Nucl.Phys.}
  {\bfseries B311} (1988) 46}.

\bibitem{Papageorgiou:2009zc}
G.~Papageorgiou and B.~J. Schroers, ``{A Chern-Simons approach to Galilean
  quantum gravity in 2+1 dimensions},''
  \href{http://dx.doi.org/10.1088/1126-6708/2009/11/009}{{\em JHEP} {\bfseries
  11} (2009) 009},
\href{http://arxiv.org/abs/0907.2880}{{\ttfamily arXiv:0907.2880 [hep-th]}}.

\bibitem{Papageorgiou:2010ud}
G.~Papageorgiou and B.~J. Schroers, ``{Galilean quantum gravity with
  cosmological constant and the extended $q$-Heisenberg algebra},''
  \href{http://dx.doi.org/10.1007/JHEP11(2010)020}{{\em JHEP} {\bfseries 11}
  (2010) 020},
\href{http://arxiv.org/abs/1008.0279}{{\ttfamily arXiv:1008.0279 [hep-th]}}.

\bibitem{Hartong:2016yrf}
J.~Hartong, Y.~Lei, and N.~A. Obers, ``{Nonrelativistic Chern-Simons theories
  and three-dimensional Hořava-Lifshitz gravity},''
  \href{http://dx.doi.org/10.1103/PhysRevD.94.065027}{{\em Phys. Rev.}
  {\bfseries D94} no.~6, (2016) 065027},
\href{http://arxiv.org/abs/1604.08054}{{\ttfamily arXiv:1604.08054 [hep-th]}}.

\bibitem{Bergshoeff:2016lwr}
E.~A. Bergshoeff and J.~Rosseel, ``{Three-Dimensional Extended Bargmann
  Supergravity},'' \href{http://dx.doi.org/10.1103/PhysRevLett.116.251601}{{\em
  Phys. Rev. Lett.} {\bfseries 116} no.~25, (2016) 251601},
\href{http://arxiv.org/abs/1604.08042}{{\ttfamily arXiv:1604.08042 [hep-th]}}.

\bibitem{Hartong:2017bwq}
J.~Hartong, Y.~Lei, N.~A. Obers, and G.~Oling, ``{Zooming in on
  AdS$_{3}$/CFT$_{2}$ near a BPS bound},''
  \href{http://dx.doi.org/10.1007/JHEP05(2018)016}{{\em JHEP} {\bfseries 05}
  (2018) 016},
\href{http://arxiv.org/abs/1712.05794}{{\ttfamily arXiv:1712.05794 [hep-th]}}.

\bibitem{Matulich:2019cdo}
J.~Matulich, S.~Prohazka, and J.~Salzer, ``{Limits of three-dimensional gravity
  and metric kinematical Lie algebras in any dimension},''
  \href{http://dx.doi.org/10.1007/JHEP07(2019)118}{{\em JHEP} {\bfseries 07}
  (2019) 118},
\href{http://arxiv.org/abs/1903.09165}{{\ttfamily arXiv:1903.09165 [hep-th]}}.

\bibitem{Bergshoeff:2016soe}
E.~Bergshoeff, D.~Grumiller, S.~Prohazka, and J.~Rosseel, ``{Three-dimensional
  Spin-3 Theories Based on General Kinematical Algebras},''
  \href{http://dx.doi.org/10.1007/JHEP01(2017)114}{{\em JHEP} {\bfseries 01}
  (2017) 114},
\href{http://arxiv.org/abs/1612.02277}{{\ttfamily arXiv:1612.02277 [hep-th]}}.

\bibitem{Gomis:2020wxp}
J.~Gomis, D.~Hidalgo, and P.~Salgado-Rebolledo, ``{Non-relativistic and
  Carrollian limits of Jackiw-Teitelboim gravity},''
  \href{http://arxiv.org/abs/2011.15053}{{\ttfamily arXiv:2011.15053
  [hep-th]}}.

\bibitem{Isler:1989hq}
K.~Isler and C.~Trugenberger, ``{A Gauge Theory of Two-dimensional Quantum
  Gravity},'' \href{http://dx.doi.org/10.1103/PhysRevLett.63.834}{{\em Phys.
  Rev. Lett.} {\bfseries 63} (1989) 834}.

\bibitem{Chamseddine:1989yz}
A.~H. Chamseddine and D.~Wyler, ``{Gauge Theory of Topological Gravity in
  (1+1)-Dimensions},''
  \href{http://dx.doi.org/10.1016/0370-2693(89)90528-5}{{\em Phys. Lett. B}
  {\bfseries 228} (1989) 75--78}.

\bibitem{Salzer:2018zlv}
J.~Salzer, {\em {Asymptotic dynamics of two-dimensional dilaton gravity}}.
\newblock PhD thesis, Vienna, Tech. U., 2018.
\newblock \url{https://inspirehep.net/files/abdb401a4bd06263c705f7772d5afba5}.

\bibitem{Birmingham:1991ty}
D.~Birmingham, M.~Blau, M.~Rakowski, and G.~Thompson, ``{Topological field
  theory},''
\href{http://dx.doi.org/10.1016/0370-1573(91)90117-5}{{\em Phys. Rept.}
  {\bfseries 209} (1991) 129--340}.

\bibitem{Medina1985}
A.~Medina and P.~Revoy, ``Alg\`{e}bres de lie et produit scalaire invariant,''
  {\em Annales scientifiques de l'\'{E}cole Normale Sup\'{e}rieure} {\bfseries
  18} no.~3, (1985) 553--561. \url{http://eudml.org/doc/82165}.

\bibitem{FigueroaO'Farrill:1995cy}
J.~M. Figueroa-O'Farrill and S.~Stanciu, ``{On the structure of symmetric
  selfdual Lie algebras},'' \href{http://dx.doi.org/10.1063/1.531620}{{\em J.
  Math. Phys.} {\bfseries 37} (1996) 4121--4134},
\href{http://arxiv.org/abs/hep-th/9506152}{{\ttfamily arXiv:hep-th/9506152
  [hep-th]}}.

\bibitem{Bacry:1968zf}
H.~Bacry and J.~Levy-Leblond, ``{Possible kinematics},''
\href{http://dx.doi.org/10.1063/1.1664490}{{\em J. Math. Phys.} {\bfseries 9}
  (1968) 1605--1614}.

\bibitem{Figueroa-OFarrill:2018ilb}
J.~Figueroa-O'Farrill and S.~Prohazka, ``{Spatially isotropic homogeneous
  spacetimes},'' \href{http://dx.doi.org/10.1007/JHEP01(2019)229}{{\em JHEP}
  {\bfseries 01} (2019) 229},
\href{http://arxiv.org/abs/1809.01224}{{\ttfamily arXiv:1809.01224 [hep-th]}}.

\bibitem{Hartong:2015zia}
J.~Hartong and N.~A. Obers, ``{Hořava-Lifshitz gravity from dynamical
  Newton-Cartan geometry},''
  \href{http://dx.doi.org/10.1007/JHEP07(2015)155}{{\em JHEP} {\bfseries 07}
  (2015) 155},
\href{http://arxiv.org/abs/1504.07461}{{\ttfamily arXiv:1504.07461 [hep-th]}}.

\bibitem{Bekaert:2015xua}
X.~Bekaert and K.~Morand, ``{Connections and dynamical trajectories in
  generalised Newton-Cartan gravity II. An ambient perspective},''
  \href{http://dx.doi.org/10.1063/1.5030328}{{\em J. Math. Phys.} {\bfseries
  59} no.~7, (2018) 072503},
\href{http://arxiv.org/abs/1505.03739}{{\ttfamily arXiv:1505.03739 [hep-th]}}.

\bibitem{Hartong:2015xda}
J.~Hartong, ``{Gauging the Carroll Algebra and Ultra-Relativistic Gravity},''
  \href{http://dx.doi.org/10.1007/JHEP08(2015)069}{{\em JHEP} {\bfseries 08}
  (2015) 069},
\href{http://arxiv.org/abs/1505.05011}{{\ttfamily arXiv:1505.05011 [hep-th]}}.

\bibitem{Hartong:2015usd}
J.~Hartong, ``{Holographic Reconstruction of 3D Flat Space-Time},''
  \href{http://dx.doi.org/10.1007/JHEP10(2016)104}{{\em JHEP} {\bfseries 10}
  (2016) 104},
\href{http://arxiv.org/abs/1511.01387}{{\ttfamily arXiv:1511.01387 [hep-th]}}.

\bibitem{Jensen:2017tnb}
K.~Jensen, ``{Locality and anomalies in warped conformal field theory},''
  \href{http://dx.doi.org/10.1007/JHEP12(2017)111}{{\em JHEP} {\bfseries 12}
  (2017) 111}, \href{http://arxiv.org/abs/1710.11626}{{\ttfamily
  arXiv:1710.11626 [hep-th]}}.

\bibitem{Figueroa-OFarrill:2020gpr}
J.~Figueroa-O'Farrill, ``{On the intrinsic torsion of spacetime structures},''
  \href{http://arxiv.org/abs/2009.01948}{{\ttfamily arXiv:2009.01948
  [hep-th]}}.

\bibitem{Bergshoeff:2017btm}
E.~Bergshoeff, J.~Gomis, B.~Rollier, J.~Rosseel, and T.~ter Veldhuis,
  ``{Carroll versus Galilei Gravity},''
  \href{http://dx.doi.org/10.1007/JHEP03(2017)165}{{\em JHEP} {\bfseries 03}
  (2017) 165},
\href{http://arxiv.org/abs/1701.06156}{{\ttfamily arXiv:1701.06156 [hep-th]}}.

\bibitem{Andringa:2010it}
R.~Andringa, E.~Bergshoeff, S.~Panda, and M.~de~Roo, ``{Newtonian Gravity and
  the Bargmann Algebra},''
  \href{http://dx.doi.org/10.1088/0264-9381/28/10/105011}{{\em Class. Quant.
  Grav.} {\bfseries 28} (2011) 105011},
  \href{http://arxiv.org/abs/1011.1145}{{\ttfamily arXiv:1011.1145 [hep-th]}}.

\bibitem{Grumiller:2006rc}
D.~Grumiller and R.~Meyer, ``{Ramifications of lineland},'' {\em Turk. J.
  Phys.} {\bfseries 30} (2006) 349--378,
  \href{http://arxiv.org/abs/hep-th/0604049}{{\ttfamily arXiv:hep-th/0604049}}.

\bibitem{Cangemi:1992bj}
D.~Cangemi and R.~Jackiw, ``{Gauge invariant formulations of lineal gravity},''
  \href{http://dx.doi.org/10.1103/PhysRevLett.69.233}{{\em Phys. Rev. Lett.}
  {\bfseries 69} (1992) 233--236},
\href{http://arxiv.org/abs/hep-th/9203056}{{\ttfamily arXiv:hep-th/9203056
  [hep-th]}}.

\bibitem{Verlinde:1991rf}
H.~L. Verlinde, ``{Black holes and strings in two-dimensions},'' in {\em {6th
  Marcel Grossmann Meeting on General Relativity (MG6)}}, pp.~178--207.
\newblock 12, 1991.

\bibitem{Jackiw:1992bw}
R.~Jackiw, ``{Gauge theories for gravity on a line},''
  \href{http://dx.doi.org/10.1007/BF01017075}{{\em Theor. Math. Phys.}
  {\bfseries 92} (1992) 979--987},
  \href{http://arxiv.org/abs/hep-th/9206093}{{\ttfamily arXiv:hep-th/9206093
  [hep-th]}}.
[,197(1992)].

\bibitem{Schrader:1972zd}
R.~Schrader, ``{The maxwell group and the quantum theory of particles in
  classical homogeneous electromagnetic fields},''
\href{http://dx.doi.org/10.1002/prop.19720201202}{{\em Fortsch. Phys.}
  {\bfseries 20} (1972) 701--734}.

\bibitem{Bacry:1970ye}
H.~Bacry, P.~Combe, and J.~L. Richard, ``{Group-theoretical analysis of
  elementary particles in an external electromagnetic field. 1. the
  relativistic particle in a constant and uniform field},''
\href{http://dx.doi.org/10.1007/BF02725178}{{\em Nuovo Cim.} {\bfseries A67}
  (1970) 267--299}.

\bibitem{Bianchi}
L.~Bianchi, ``Sugli spazi a tre dimensioni che ammettono un gruppo continuo di
  movimenti,'' {\em Memorie di Matematica e di Fisica della Societa Italiana
  delle Scienze, Serie Terza,} {\bfseries Tomo XI} (1898) 267--352.

\bibitem{MR1900159}
L.~Bianchi, ``On the three-dimensional spaces which admit a continuous group of
  motions,'' \href{http://dx.doi.org/10.1023/A:1015357132699}{{\em Gen.
  Relativity Gravitation} {\bfseries 33} no.~12, (2001) 2171--2253}.
  \url{https://doi-org.ezproxy.is.ed.ac.uk/10.1023/ A:1015357132699}.
  Translated from the Italian by R. Jantzen.

\bibitem{Witten:1988hf}
E.~Witten, ``{Quantum Field Theory and the Jones Polynomial},''
\href{http://dx.doi.org/10.1007/BF01217730}{{\em Commun.Math.Phys.} {\bfseries
  121} (1989) 351--399}.

\bibitem{Elitzur:1989nr}
S.~Elitzur, G.~W. Moore, A.~Schwimmer, and N.~Seiberg, ``{Remarks on the
  Canonical Quantization of the Chern-Simons-Witten Theory},''
\href{http://dx.doi.org/10.1016/0550-3213(89)90436-7}{{\em Nucl. Phys.}
  {\bfseries B326} (1989) 108}.

\bibitem{Gonzalez:2018enk}
H.~A. González, D.~Grumiller, and J.~Salzer, ``{Towards a bulk description of
  higher spin SYK},'' \href{http://dx.doi.org/10.1007/JHEP05(2018)083}{{\em
  JHEP} {\bfseries 05} (2018) 083},
\href{http://arxiv.org/abs/1802.01562}{{\ttfamily arXiv:1802.01562 [hep-th]}}.

\bibitem{Kapec:2019ecr}
D.~Kapec, R.~Mahajan, and D.~Stanford, ``{Matrix ensembles with global
  symmetries and \textquoteright{}t Hooft anomalies from 2d gauge theory},''
  \href{http://dx.doi.org/10.1007/JHEP04(2020)186}{{\em JHEP} {\bfseries 04}
  (2020) 186}, \href{http://arxiv.org/abs/1912.12285}{{\ttfamily
  arXiv:1912.12285 [hep-th]}}.

\bibitem{Cotler:2018zff}
J.~Cotler and K.~Jensen, ``{A theory of reparameterizations for AdS$_3$
  gravity},'' \href{http://dx.doi.org/10.1007/JHEP02(2019)079}{{\em JHEP}
  {\bfseries 02} (2019) 079}, \href{http://arxiv.org/abs/1808.03263}{{\ttfamily
  arXiv:1808.03263 [hep-th]}}.

\bibitem{Merbis:2019wgk}
W.~Merbis and M.~Riegler, ``{Geometric actions and flat space holography},''
  \href{http://dx.doi.org/10.1007/JHEP02(2020)125}{{\em JHEP} {\bfseries 02}
  (2020) 125}, \href{http://arxiv.org/abs/1912.08207}{{\ttfamily
  arXiv:1912.08207 [hep-th]}}.

\bibitem{Brown:1986nw}
J.~D. Brown and M.~Henneaux, ``{Central Charges in the Canonical Realization of
  Asymptotic Symmetries: An Example from Three-Dimensional Gravity},''
\href{http://dx.doi.org/10.1007/BF01211590}{{\em Commun.Math.Phys.} {\bfseries
  104} (1986) 207--226}.

\bibitem{Compere:2013bya}
G.~Comp\`ere, W.~Song, and A.~Strominger, ``{New Boundary Conditions for
  AdS3},'' \href{http://dx.doi.org/10.1007/JHEP05(2013)152}{{\em JHEP}
  {\bfseries 05} (2013) 152}, \href{http://arxiv.org/abs/1303.2662}{{\ttfamily
  arXiv:1303.2662 [hep-th]}}.

\bibitem{Troessaert:2013fma}
C.~Troessaert, ``{Enhanced asymptotic symmetry algebra of $AdS$$_{3}$},''
  \href{http://dx.doi.org/10.1007/JHEP08(2013)044}{{\em JHEP} {\bfseries 08}
  (2013) 044}, \href{http://arxiv.org/abs/1303.3296}{{\ttfamily arXiv:1303.3296
  [hep-th]}}.

\bibitem{Grumiller:2016pqb}
D.~Grumiller and M.~Riegler, ``{Most general AdS$_{3}$ boundary conditions},''
  \href{http://dx.doi.org/10.1007/JHEP10(2016)023}{{\em JHEP} {\bfseries 10}
  (2016) 023},
\href{http://arxiv.org/abs/1608.01308}{{\ttfamily arXiv:1608.01308 [hep-th]}}.

\bibitem{Coussaert:1995zp}
O.~Coussaert, M.~Henneaux, and P.~van Driel, ``{The Asymptotic dynamics of
  three-dimensional Einstein gravity with a negative cosmological constant},''
  \href{http://dx.doi.org/10.1088/0264-9381/12/12/012}{{\em Class.Quant.Grav.}
  {\bfseries 12} (1995) 2961--2966},
\href{http://arxiv.org/abs/gr-qc/9506019}{{\ttfamily arXiv:gr-qc/9506019
  [gr-qc]}}.

\bibitem{Barnich:2006av}
G.~Barnich and G.~Compere, ``{Classical central extension for asymptotic
  symmetries at null infinity in three spacetime dimensions},''
  \href{http://dx.doi.org/10.1088/0264-9381/24/5/F01,
  10.1088/0264-9381/24/11/C01}{{\em Class. Quant. Grav.} {\bfseries 24} (2007)
  F15--F23},
\href{http://arxiv.org/abs/gr-qc/0610130}{{\ttfamily arXiv:gr-qc/0610130
  [gr-qc]}}.

\bibitem{Feher:1992ed}
L.~Feher, L.~O'Raifeartaigh, P.~Ruelle, I.~Tsutsui, and A.~Wipf, ``{On the
  general structure of Hamiltonian reductions of the WZNW theory},''
\href{http://arxiv.org/abs/hep-th/9112068}{{\ttfamily arXiv:hep-th/9112068
  [hep-th]}}.

\bibitem{Henneaux:1992ig}
M.~Henneaux and C.~Teitelboim, {\em {Quantization of gauge systems}}.
\newblock Princeton University Press, Princeton, NJ,
1992.
\newblock

\bibitem{Grumiller:2017qao}
D.~Grumiller, R.~McNees, J.~Salzer, C.~Valcárcel, and D.~Vassilevich,
  ``{Menagerie of AdS$_{2}$ boundary conditions},''
  \href{http://dx.doi.org/10.1007/JHEP10(2017)203}{{\em JHEP} {\bfseries 10}
  (2017) 203},
\href{http://arxiv.org/abs/1708.08471}{{\ttfamily arXiv:1708.08471 [hep-th]}}.

\bibitem{Cotler:2019nbi}
J.~Cotler, K.~Jensen, and A.~Maloney, ``{Low-dimensional de Sitter quantum
  gravity},'' \href{http://dx.doi.org/10.1007/JHEP06(2020)048}{{\em JHEP}
  {\bfseries 06} (2020) 048}, \href{http://arxiv.org/abs/1905.03780}{{\ttfamily
  arXiv:1905.03780 [hep-th]}}.

\bibitem{Carlip:2016lnw}
S.~Carlip, ``{The dynamics of supertranslations and superrotations in 2 + 1
  dimensions},'' \href{http://dx.doi.org/10.1088/1361-6382/aa9809}{{\em Class.
  Quant. Grav.} {\bfseries 35} no.~1, (2018) 014001},
  \href{http://arxiv.org/abs/1608.05088}{{\ttfamily arXiv:1608.05088 [gr-qc]}}.

\bibitem{Nguyen:2020hot}
K.~Nguyen and J.~Salzer, ``{The Effective Action of Superrotation Modes},''
  \href{http://arxiv.org/abs/2008.03321}{{\ttfamily arXiv:2008.03321
  [hep-th]}}.

\bibitem{Afshar:2015wjm}
H.~Afshar, S.~Detournay, D.~Grumiller, and B.~Oblak, ``{Near-Horizon Geometry
  and Warped Conformal Symmetry},''
  \href{http://dx.doi.org/10.1007/JHEP03(2016)187}{{\em JHEP} {\bfseries 03}
  (2016) 187},
\href{http://arxiv.org/abs/1512.08233}{{\ttfamily arXiv:1512.08233 [hep-th]}}.

\bibitem{Afshar:2019tvp}
H.~R. Afshar, ``{Warped Schwarzian theory},''
  \href{http://dx.doi.org/10.1007/JHEP02(2020)126}{{\em JHEP} {\bfseries 02}
  (2020) 126}, \href{http://arxiv.org/abs/1908.08089}{{\ttfamily
  arXiv:1908.08089 [hep-th]}}.

\bibitem{Afshar:2019axx}
H.~Afshar, H.~A. González, D.~Grumiller, and D.~Vassilevich, ``{Flat space
  holography and the complex Sachdev-Ye-Kitaev model},''
  \href{http://dx.doi.org/10.1103/PhysRevD.101.086024}{{\em Phys. Rev. D}
  {\bfseries 101} no.~8, (2020) 086024},
  \href{http://arxiv.org/abs/1911.05739}{{\ttfamily arXiv:1911.05739
  [hep-th]}}.

\bibitem{Duval:2014uoa}
C.~Duval, G.~Gibbons, P.~Horvathy, and P.~Zhang, ``{Carroll versus Newton and
  Galilei: two dual non-Einsteinian concepts of time},''
  \href{http://dx.doi.org/10.1088/0264-9381/31/8/085016}{{\em Class. Quant.
  Grav.} {\bfseries 31} (2014) 085016},
  \href{http://arxiv.org/abs/1402.0657}{{\ttfamily arXiv:1402.0657 [gr-qc]}}.

\bibitem{Bagchi:2019clu}
A.~Bagchi, R.~Basu, A.~Mehra, and P.~Nandi, ``{Field Theories on Null
  Manifolds},'' \href{http://dx.doi.org/10.1007/JHEP02(2020)141}{{\em JHEP}
  {\bfseries 02} (2020) 141}, \href{http://arxiv.org/abs/1912.09388}{{\ttfamily
  arXiv:1912.09388 [hep-th]}}.

\bibitem{Ozdemir:2019orp}
N.~Ozdemir, M.~Ozkan, O.~Tunca, and U.~Zorba, ``{Three-Dimensional Extended
  Newtonian (Super)Gravity},''
  \href{http://dx.doi.org/10.1007/JHEP05(2019)130}{{\em JHEP} {\bfseries 05}
  (2019) 130}, \href{http://arxiv.org/abs/1903.09377}{{\ttfamily
  arXiv:1903.09377 [hep-th]}}.

\bibitem{Ravera:2019ize}
L.~Ravera, ``{AdS Carroll Chern-Simons supergravity in 2 + 1 dimensions and its
  flat limit},'' \href{http://dx.doi.org/10.1016/j.physletb.2019.06.026}{{\em
  Phys. Lett. B} {\bfseries 795} (2019) 331--338},
  \href{http://arxiv.org/abs/1905.00766}{{\ttfamily arXiv:1905.00766
  [hep-th]}}.

\bibitem{Concha:2019lhn}
P.~Concha and E.~Rodr\'\i{}guez, ``{Non-Relativistic Gravity Theory based on an
  Enlargement of the Extended Bargmann Algebra},''
  \href{http://dx.doi.org/10.1007/JHEP07(2019)085}{{\em JHEP} {\bfseries 07}
  (2019) 085}, \href{http://arxiv.org/abs/1906.00086}{{\ttfamily
  arXiv:1906.00086 [hep-th]}}.

\bibitem{Gomis:2019nih}
J.~Gomis, A.~Kleinschmidt, J.~Palmkvist, and P.~Salgado-Rebolledo,
  ``{Newton-Hooke/Carrollian expansions of (A)dS and Chern-Simons gravity},''
  \href{http://dx.doi.org/10.1007/JHEP02(2020)009}{{\em JHEP} {\bfseries 02}
  (2020) 009}, \href{http://arxiv.org/abs/1912.07564}{{\ttfamily
  arXiv:1912.07564 [hep-th]}}.

\bibitem{Mohammedi:1993rg}
N.~Mohammedi, ``{On bosonic and supersymmetric current algebras for
  nonsemisimple groups},''
  \href{http://dx.doi.org/10.1016/0370-2693(94)90027-2}{{\em Phys. Lett.}
  {\bfseries B325} (1994) 371--376},
\href{http://arxiv.org/abs/hep-th/9312182}{{\ttfamily arXiv:hep-th/9312182
  [hep-th]}}.

\bibitem{FigueroaO'Farrill:1994hx}
J.~M. Figueroa-O'Farrill and S.~Stanciu, ``{Nonsemisimple Sugawara
  constructions},'' \href{http://dx.doi.org/10.1016/0370-2693(94)91525-3}{{\em
  Phys. Lett.} {\bfseries B327} (1994) 40--46},
\href{http://arxiv.org/abs/hep-th/9402035}{{\ttfamily arXiv:hep-th/9402035
  [hep-th]}}.

\bibitem{Chaturvedi:2020jyy}
P.~Chaturvedi, I.~Papadimitriou, W.~Song, and B.~Yu, ``{AdS$_3$ gravity and the
  complex SYK models},'' \href{http://arxiv.org/abs/2011.10001}{{\ttfamily
  arXiv:2011.10001 [hep-th]}}.

\bibitem{Grumiller:2017sjh}
D.~Grumiller, W.~Merbis, and M.~Riegler, ``{Most general flat space boundary
  conditions in three-dimensional Einstein gravity},''
  \href{http://dx.doi.org/10.1088/1361-6382/aa8004}{{\em Class. Quant. Grav.}
  {\bfseries 34} (2017) 184001},
\href{http://arxiv.org/abs/1704.07419}{{\ttfamily arXiv:1704.07419 [hep-th]}}.

\bibitem{Strominger:2017zoo}
A.~Strominger, ``{Lectures on the Infrared Structure of Gravity and Gauge
  Theory},''
\href{http://arxiv.org/abs/1703.05448}{{\ttfamily arXiv:1703.05448 [hep-th]}}.

\bibitem{Guica:2017lia}
M.~Guica, ``{An integrable Lorentz-breaking deformation of two-dimensional
  CFTs},'' \href{http://dx.doi.org/10.21468/SciPostPhys.5.5.048}{{\em SciPost
  Phys.} {\bfseries 5} no.~5, (2018) 048},
  \href{http://arxiv.org/abs/1710.08415}{{\ttfamily arXiv:1710.08415
  [hep-th]}}.

\bibitem{Gross:2019ach}
D.~J. Gross, J.~Kruthoff, A.~Rolph, and E.~Shaghoulian, ``{$T\overline{T}$ in
  AdS$_2$ and Quantum Mechanics},''
  \href{http://dx.doi.org/10.1103/PhysRevD.101.026011}{{\em Phys. Rev. D}
  {\bfseries 101} no.~2, (2020) 026011},
  \href{http://arxiv.org/abs/1907.04873}{{\ttfamily arXiv:1907.04873
  [hep-th]}}.

\bibitem{Gross:2019uxi}
D.~J. Gross, J.~Kruthoff, A.~Rolph, and E.~Shaghoulian, ``{Hamiltonian
  deformations in quantum mechanics, $T\bar T$, and the SYK model},''
  \href{http://dx.doi.org/10.1103/PhysRevD.102.046019}{{\em Phys. Rev. D}
  {\bfseries 102} no.~4, (2020) 046019},
  \href{http://arxiv.org/abs/1912.06132}{{\ttfamily arXiv:1912.06132
  [hep-th]}}.

\bibitem{Grumiller:2020fbb}
D.~Grumiller and R.~McNees, ``{Universal flow equations and chaos bound
  saturation in 2d dilaton gravity},''
  \href{http://arxiv.org/abs/2007.03673}{{\ttfamily arXiv:2007.03673
  [hep-th]}}.

\bibitem{Cardenas:2018krd}
M.~C\'ardenas, O.~Fuentealba, H.~A. Gonz\'alez, D.~Grumiller, C.~Valc\'arcel,
  and D.~Vassilevich, ``{Boundary theories for dilaton supergravity in 2D},''
  \href{http://dx.doi.org/10.1007/JHEP11(2018)077}{{\em JHEP} {\bfseries 11}
  (2018) 077}, \href{http://arxiv.org/abs/1809.07208}{{\ttfamily
  arXiv:1809.07208 [hep-th]}}.

\bibitem{Alkalaev:2020kut}
K.~Alkalaev and X.~Bekaert, ``{On BF-type higher-spin actions in two
  dimensions},'' \href{http://dx.doi.org/10.1007/JHEP05(2020)158}{{\em JHEP}
  {\bfseries 05} (2020) 158}, \href{http://arxiv.org/abs/2002.02387}{{\ttfamily
  arXiv:2002.02387 [hep-th]}}.

\bibitem{Barnich:1993vg}
G.~Barnich and M.~Henneaux, ``{Consistent couplings between fields with a gauge
  freedom and deformations of the master equation},''
  \href{http://dx.doi.org/10.1016/0370-2693(93)90544-R}{{\em Phys. Lett. B}
  {\bfseries 311} (1993) 123--129},
  \href{http://arxiv.org/abs/hep-th/9304057}{{\ttfamily arXiv:hep-th/9304057}}.

\bibitem{Ikeda:1993fh}
N.~Ikeda, ``{Two-dimensional gravity and nonlinear gauge theory},''
  \href{http://dx.doi.org/10.1006/aphy.1994.1104}{{\em Annals Phys.} {\bfseries
  235} (1994) 435--464}, \href{http://arxiv.org/abs/hep-th/9312059}{{\ttfamily
  arXiv:hep-th/9312059}}.

\bibitem{Schaller:1994es}
P.~Schaller and T.~Strobl, ``{Poisson structure induced (topological) field
  theories},'' \href{http://dx.doi.org/10.1142/S0217732394002951}{{\em Mod.
  Phys. Lett. A} {\bfseries 9} (1994) 3129--3136},
  \href{http://arxiv.org/abs/hep-th/9405110}{{\ttfamily arXiv:hep-th/9405110}}.

\bibitem{Ovando:Review}
G.~P. {Ovando}, ``{Lie algebras with ad-invariant metrics. A survey},''
  \href{http://arxiv.org/abs/1512.03997}{{\ttfamily arXiv:1512.03997
  [math.DG]}}.

\bibitem{Medina}
A.~Medina, ``Groupes de {L}ie munis de m\'{e}triques bi-invariantes,''
  \href{http://dx.doi.org/10.2748/tmj/1178228586}{{\em Tohoku Math. J. (2)}
  {\bfseries 37} no.~4, (1985) 405--421}.
  \url{https://doi.org/10.2748/tmj/1178228586}.

\bibitem{Figueroa-OFarrill:2003fkz}
J.~M. Figueroa-O'Farrill, ``{On parallelizable NS NS backgrounds},''
  \href{http://dx.doi.org/10.1088/0264-9381/20/15/304}{{\em Class. Quant.
  Grav.} {\bfseries 20} (2003) 3327--3340},
  \href{http://arxiv.org/abs/hep-th/0305079}{{\ttfamily arXiv:hep-th/0305079}}.

\bibitem{KathOlb:2004}
I.~Kath and M.~Olbrich, ``Metric {L}ie algebras with maximal isotropic
  centre,'' \href{http://dx.doi.org/10.1007/s00209-003-0575-2}{{\em Math. Z.}
  {\bfseries 246} no.~1-2, (2004) 23--53}.
  \url{https://doi.org/10.1007/s00209-003-0575-2}.

\bibitem{KathOlb:2006}
I.~Kath and M.~Olbrich, ``Metric {L}ie algebras and quadratic extensions,''
  \href{http://dx.doi.org/10.1007/s00031-005-1106-5}{{\em Transform. Groups}
  {\bfseries 11} no.~1, (2006) 87--131}.
  \url{https://doi.org/10.1007/s00031-005-1106-5}.

\bibitem{Hansen:2018ofj}
D.~Hansen, J.~Hartong, and N.~A. Obers, ``{Action Principle for Newtonian
  Gravity},'' \href{http://dx.doi.org/10.1103/PhysRevLett.122.061106}{{\em
  Phys. Rev. Lett.} {\bfseries 122} no.~6, (2019) 061106},
\href{http://arxiv.org/abs/1807.04765}{{\ttfamily arXiv:1807.04765 [hep-th]}}.

\bibitem{deAzcarraga:2002xi}
J.~A. de~Azcarraga, J.~M. Izquierdo, M.~Picon, and O.~Varela, ``{Generating Lie
  and gauge free differential (super)algebras by expanding Maurer-Cartan forms
  and Chern-Simons supergravity},''
  \href{http://dx.doi.org/10.1016/S0550-3213(03)00342-0}{{\em Nucl. Phys.}
  {\bfseries B662} (2003) 185--219},
\href{http://arxiv.org/abs/hep-th/0212347}{{\ttfamily arXiv:hep-th/0212347
  [hep-th]}}.

\bibitem{Izaurieta:2006zz}
F.~Izaurieta, E.~Rodriguez, and P.~Salgado, ``{Expanding Lie (super)algebras
  through Abelian semigroups},''
  \href{http://dx.doi.org/10.1063/1.2390659}{{\em J. Math. Phys.} {\bfseries
  47} (2006) 123512},
\href{http://arxiv.org/abs/hep-th/0606215}{{\ttfamily arXiv:hep-th/0606215
  [hep-th]}}.

\bibitem{Khasanov:2011jr}
O.~Khasanov and S.~Kuperstein, ``{(In)finite extensions of algebras from their
  Inonu-Wigner contractions},''
  \href{http://dx.doi.org/10.1088/1751-8113/44/47/475202}{{\em J. Phys. A}
  {\bfseries 44} (2011) 475202},
  \href{http://arxiv.org/abs/1103.3447}{{\ttfamily arXiv:1103.3447 [hep-th]}}.

\bibitem{Prohazka:2017pkc}
S.~Prohazka, {\em {Chern-Simons Holography: Boundary Conditions, Contractions
  and Double Extensions for a Journey Beyond Anti-de Sitter}}.
\newblock PhD thesis, Vienna, Tech. U., 2017.
\newblock
\href{http://arxiv.org/abs/1710.11110}{{\ttfamily arXiv:1710.11110 [hep-th]}}.
\newblock

\bibitem{Barducci:2019jhj}
A.~Barducci, R.~Casalbuoni, and J.~Gomis, ``{Nonrelativistic $k$-contractions
  of the coadjoint Poincar\'e algebra},''
  \href{http://dx.doi.org/10.1142/S0217751X20500098}{{\em Int. J. Mod. Phys. A}
  {\bfseries 35} no.~04, (2020) 2050009},
  \href{http://arxiv.org/abs/1910.11682}{{\ttfamily arXiv:1910.11682
  [physics.gen-ph]}}.

\bibitem{Bergshoeff:2020fiz}
E.~Bergshoeff, J.~Gomis, and P.~Salgado-Rebolledo, ``{Non-relativistic limits
  and three-dimensional coadjoint Poincare gravity},''
  \href{http://dx.doi.org/10.1098/rspa.2020.0106}{{\em Proc. Roy. Soc. Lond. A}
  {\bfseries 476} no.~2240, (2020) 20200106},
  \href{http://arxiv.org/abs/2001.11790}{{\ttfamily arXiv:2001.11790
  [hep-th]}}.

\bibitem{Barducci:2020blv}
A.~Barducci, R.~Casalbuoni, and J.~Gomis, ``{A particle model with extra
  dimensions from Coadjoint Poincar\'e Symmetry},''
  \href{http://dx.doi.org/10.1007/JHEP08(2020)092}{{\em JHEP} {\bfseries 08}
  (2020) 092}, \href{http://arxiv.org/abs/2006.11725}{{\ttfamily
  arXiv:2006.11725 [hep-th]}}.

\end{thebibliography}\endgroup

\end{document}